\documentclass{article}
\usepackage{ijcai19}

\usepackage{times}
\usepackage{soul}
\usepackage{url}
\usepackage[hidelinks]{hyperref}
\usepackage[utf8]{inputenc}
\usepackage[small]{caption}
\usepackage{graphicx}
\usepackage{amsmath}
\usepackage{amssymb}
\usepackage{bbm}
\usepackage{booktabs}
\usepackage{algorithm}
\usepackage{algorithmic}
\usepackage{multirow}
\usepackage{enumitem}
\usepackage{natbib} 
\usepackage[font=small,labelfont=bf]{caption}
\title{Toward Fine-Grained Speech Inpainting Forensics: \\
A Dataset, Method, and Metric for \\
Multi-Region Tampering Localization}

\author{
Tung Vu$^1$ \and
Yen Nguyen$^1$ \and
Hai Nguyen$^1$ \and
Cuong Pham$^1$ \and
Cong Tran$^1$\footnote{Corresponding author}\\
\affiliations
$^1$Posts and Telecommunications Institute of Technology, Hanoi, Vietnam\\
\emails
\{tung.vuson.hau, yen1422mh, namhai1810k2003\}@gmail.com,
\{cuongpv, congtt\}@ptit.edu.vn
}
\begin{document}

\maketitle

\begin{abstract}
Recent advances in voice cloning and text-to-speech synthesis have made
partial speech manipulation---where an adversary replaces a few words within an utterance to alter its meaning while preserving the speaker's
identity---an increasingly realistic threat. 
Existing audio deepfake detection benchmarks focus on utterance-level binary classification or single-region tampering, leaving a critical gap in detecting and localizing \emph{multiple} inpainted segments whose count is unknown a priori. We address this gap with three contributions.
First, we introduce \textbf{MIST} (\textbf{M}ulti-region \textbf{I}npainting \textbf{S}peech \textbf{T}ampering), a large-scale multilingual dataset spanning 6 languages with 1--3 independently inpainted word-level segments per utterance, generated via LLM-guided semantic replacement and neural voice cloning, with fake content constituting only 2--7\% of each utterance. Second, we propose \textbf{ISA} (\textbf{I}terative \textbf{S}egment \textbf{A}nalysis), a backbone-agnostic framework that performs coarse-to-fine sliding-window classification with gap-tolerant region proposal and boundary refinement to recover all tampered regions without prior knowledge of their count. Third, we define \textbf{SF1@$\tau$}, a segment-level F1 metric based on temporal IoU matching that jointly evaluates region count accuracy and
localization precision. Zero-shot evaluation reveals that partial inpainting at word granularity remains unsolved by existing deepfake detectors: utterance-level classifiers trained on fully synthesized speech assign near-zero fake probability to MIST utterances where only 2--7\% of content is manipulated. ISA consistently outperforms non-iterative baselines in this challenging setting, and the dataset, code, and evaluation toolkit are publicly released.\footnote{\url{https://huggingface.co/datasets/tung2308/MIST_SpeechInpaintingDataset}}
\end{abstract}

\section{Introduction}
\label{sec:intro}

The proliferation of neural text-to-speech (TTS) and voice conversion (VC)
technologies has given rise to increasingly sophisticated audio
deepfakes~\cite{yi2023audio_deepfake_survey}.
While fully synthesized speech has received extensive research attention,
a more insidious form of manipulation---\emph{partial speech
inpainting}---poses a uniquely dangerous threat.
In this scenario, an adversary replaces only a few carefully chosen words
within a genuine utterance, preserving the original speaker's voice
characteristics, prosody, and recording conditions for the vast majority
of the signal.
By changing as few as one to three semantically critical words, the meaning
of a statement can be drastically altered
(e.g., ``I \emph{support} this policy'' $\rightarrow$
``I \emph{oppose} this policy'') while remaining nearly imperceptible to
human listeners.
Unlike fully synthesized speech, the fake content in a partial inpainting
attack constitutes only 2--7\% of the utterance duration, making it
orders of magnitude harder to detect and precisely localize.

Existing audio deepfake detection systems have achieved remarkable progress
on the fully-synthesized speech scenario.
Utterance-level classifiers such as RawNet2~\cite{tak2021rawnet2} and
AASIST~\cite{jung2022aasist} operate directly on raw waveforms or
spectro-temporal graphs to produce a single binary real/fake decision.
Self-supervised approaches leveraging Wav2Vec~2.0~\cite{baevski2020wav2vec2}
and WavLM~\cite{chen2022wavlm} features have further pushed performance
on ASVspoof benchmarks~\cite{wang2020asvspoof2019,nautsch2021asvspoof2019,
yamagishi2021asvspoof2021acceleratingprogress}.
For partial manipulation, PartialSpoof~\cite{zhang2023partialspoof}
introduced multi-resolution countermeasures for simultaneous utterance- and
segment-level detection, while Half-Truth~\cite{yi2021halftruth}, LAV-DF~\cite{cai2022lavdf}, and LlamaPartialSpoof~\cite{luong2025llamapartialspoofllmdrivenfakespeech}
explored single-region splicing at increasing scale.
Despite these advances, all existing approaches share a common assumption:
each utterance contains \emph{at most one} contiguous tampered region,
and its presence is confirmed by a binary utterance-level label.

This assumption breaks down precisely in the most realistic and dangerous
attack scenario: an adversary who replaces \emph{multiple} scattered words
to alter the conveyed message.
Three interrelated gaps prevent existing work from addressing this threat.
\textbf{Dataset gap:} no publicly available benchmark provides utterances
with more than one independently inpainted region, multilingual coverage,
or word-level temporal annotations for each fake segment.
\textbf{Methodological gap:} even methods that attempt temporal localization
assume a fixed or known number of tampered regions; when the manipulation
count is unknown a priori, frame-level approaches produce fragmented
predictions with no principled aggregation into coherent segment
hypotheses.
\textbf{Evaluation gap:} standard metrics---utterance-level accuracy, equal
error rate (EER), or frame-level AUC---penalize neither over-segmentation
nor under-segmentation and thus fail to capture the dual challenge of
correctly \emph{counting} tampered regions and precisely
\emph{localizing} their temporal boundaries.
Our zero-shot experiments confirm the severity of the methodological gap:
state-of-the-art utterance-level deepfake classifiers assign near-zero
fake probability to utterances where only 2--7\% of content is
manipulated, regardless of the inference-time strategy applied.

To address these gaps, we make the following contributions:
\begin{enumerate}[leftmargin=*, itemsep=2pt]

    \item \textbf{MIST Dataset} (Section~\ref{sec:dataset}):
    A large-scale multilingual benchmark spanning five languages
    (Chinese, English, French, Japanese, Vietnamese) with 1--3
    independently inpainted word-level segments per utterance,
    generated via LLM-guided semantic replacement and state-of-the-art
    neural voice cloning.
    MIST contains 598k utterances (478\,h genuine, 1{,}119\,h inpainted)
    and provides precise word-level temporal annotations for every
    fake segment, making it the first dataset to systematically evaluate
    multi-region partial inpainting detection.

    \item \textbf{ISA Method} (Section~\ref{sec:method}):
    An Iterative Segment Analysis pipeline that performs coarse-to-fine
    sliding-window classification, gap-tolerant region proposal merging,
    and boundary refinement to localize all tampered regions without
    requiring prior knowledge of their count.
    ISA is backbone-agnostic and introduces no additional trainable
    parameters beyond the underlying classifier.

    \item \textbf{SF1@$\tau$ Metric} (Section~\ref{sec:metric}):
    A segment-level F1 score based on temporal IoU matching that jointly
    evaluates region count accuracy and localization precision in a single
    interpretable number, complemented by Count Accuracy (CA) to
    disentangle counting errors from boundary errors.

\end{enumerate}

\noindent
Experiments demonstrate that ISA consistently outperforms frame-level and
single-window baselines even in the zero-shot regime, and our analysis
establishes the first quantitative evidence that partial inpainting at
word granularity is an open, unsolved problem for the audio forensics
community.
The MIST dataset, ISA codebase, and SF1@$\tau$ evaluation toolkit are
publicly released to accelerate progress on this
challenge.\footnote{\url{https://huggingface.co/datasets/tung2308/MIST_SpeechInpaintingDataset}}
\section{Related Work}
\label{sec:related}

\subsection{Audio Deepfake Detection}

Audio deepfake detection has been extensively studied in the context of automatic speaker verification (ASV) spoofing countermeasures.
The ASVspoof challenge series~\cite{wang2020asvspoof2019,nautsch2021asvspoof2019,yamagishi2021asvspoof2021} has driven progress through standardized benchmarks covering TTS, VC, and replay attacks.
State-of-the-art systems include RawNet2~\cite{tak2021rawnet2}, which operates on raw waveforms; AASIST~\cite{jung2022aasist}, which combines spectral and temporal graph attention; and self-supervised approaches leveraging Wav2Vec 2.0~\cite{baevski2020wav2vec2} and WavLM~\cite{chen2022wavlm} features for spoofing detection~\cite{tak2022wav2vec_spoofing}.
However, these methods produce \emph{utterance-level} binary decisions and are not designed to localize tampered regions within partially manipulated speech.

\subsection{Partial Speech Manipulation and Datasets}

Partial manipulation---where only a segment of an utterance is replaced---has gained attention as a realistic attack vector.
The PartialSpoof dataset~\cite{zhang2023partialspoof,zhang2021partialspoof_initial} introduced utterances with a single contiguous spliced region and segment-level labels.
The Half-Truth dataset~\cite{yi2021halfthruth} combined real and synthetic segments at utterance boundaries.
More recently, LAV-DF~\cite{cai2022lavdf} used a rule-based system to replace words with antonyms, and AV-Deepfake1M~\cite{cai2023avdeepfake1m} employed ChatGPT to alter sentences.
LlamaPartialSpoof~\cite{luong2024llamapartialspoof} demonstrated LLM-driven partial manipulation at scale.
Negroni et al.~\cite{negroni2024psynd} analyzed the impact of splicing artifacts in partially fake speech.
While these works represent important steps, they are limited to \emph{single-region} tampering or use only one or two TTS models.
In contrast, real-world adversaries may replace \emph{multiple short words} scattered across an utterance, a scenario not covered by existing datasets.

\subsection{Tampering Localization}

Temporal localization of audio tampering has been approached through frame-level classification~\cite{zhang2023partialspoof}, boundary detection, and attention-based methods.
The PartialSpoof work~\cite{zhang2023partialspoof} proposed multi-resolution countermeasures for simultaneous utterance- and segment-level detection.
Most methods assume a known or fixed number of tampered regions.
When the number of manipulated segments is unknown, frame-level approaches suffer from fragmented predictions and lack a principled mechanism to aggregate frame decisions into coherent segment-level hypotheses.
Our proposed ISA method addresses this limitation through iterative region proposal and refinement.

\section{MIST Dataset}
\label{sec:dataset}

We introduce \textbf{MIST} (\textbf{M}ulti-region \textbf{I}npainting \textbf{S}peech
\textbf{T}ampering),
a large-scale multilingual dataset for benchmarking multi-region audio inpainting
detection and localization.
Unlike existing datasets that are predominantly monolingual and limited to
single-region tampering, MIST spans \textbf{six languages}---English~(EN),
French~(FR), German~(DE), Italian~(IT), Spanish~(ES), and Vietnamese~(VI)---%
with up to three independently inpainted word-level segments per utterance
and precise word-level temporal annotations.

\textbf{Source corpora.}
The genuine speech in MIST is drawn from two complementary open-source corpora.
For English, French, German, Italian, and Spanish we use the
\textbf{Multilingual LibriSpeech} (MLS) collection~\cite{pratap2020mls},
available at \url{https://huggingface.co/datasets/openslr/librispeech_asr},
which provides audiobook recordings with high-quality forced-alignment
word-level timestamps across all five languages.
For Vietnamese, we draw from the \textbf{LEMAS-Dataset}~\cite{zhao2026lemas},
the largest open-source multilingual speech corpus with word-level timestamps,
covering over 150{,}000 hours across 10 major languages.
We select approximately 30\,GB of speech per language from each collection,
leveraging their high-quality forced-alignment timestamps as the foundation
for our word selection and splicing pipeline.
The availability of precise word boundaries eliminates the need for a
separate forced-alignment step and ensures accurate temporal annotations
for tampered regions.

\subsection{Dataset Design}
\label{sec:threat_model}

Our dataset is motivated by a practical disinformation scenario: an adversary
who has access to a recording of a target speaker aims to alter the meaning of
an utterance by replacing a small number of semantically critical words, while
preserving the speaker's identity, prosody, and recording conditions for the
vast majority of the signal.
This \emph{partial inpainting} attack is particularly dangerous because
(i)~it leaves most of the audio untouched, making it difficult for both
human listeners and automated detectors, and (ii)~it can drastically change
the conveyed message with minimal manipulation.

\textbf{Multilingual scope.}
Real-world disinformation is not confined to a single language.
To evaluate detector robustness across diverse phonological systems, we include
six typologically varied languages:
English,
French,
German,
Italian,
Spanish, and
Vietnamese.
This diversity spans three major Romance languages alongside Germanic English
and the tone-rich Vietnamese, ensuring that detection methods must generalize
beyond language-specific acoustic cues.

\textbf{Duration-aware variant strategy.}
To avoid unrealistic manipulation densities, we adopt a duration-aware
generation strategy:
\begin{itemize}[leftmargin=*, itemsep=2pt]
    \item Utterances shorter than $\theta$ seconds ($\theta\!=\!10$\,s) yield
          \textbf{2 variants}: 1-word and 2-word replacements.
    \item Utterances of $\theta$ seconds or longer yield \textbf{3 variants}:
          1-word, 2-word, and 3-word replacements.
\end{itemize}
Each variant is generated independently with different randomly selected
target words, resulting in a rich set of manipulation patterns per source
utterance.

\subsection{Generation Pipeline}
\label{sec:pipeline}

\begin{figure*}[htbp]
    \centering
    \includegraphics[width=\textwidth]{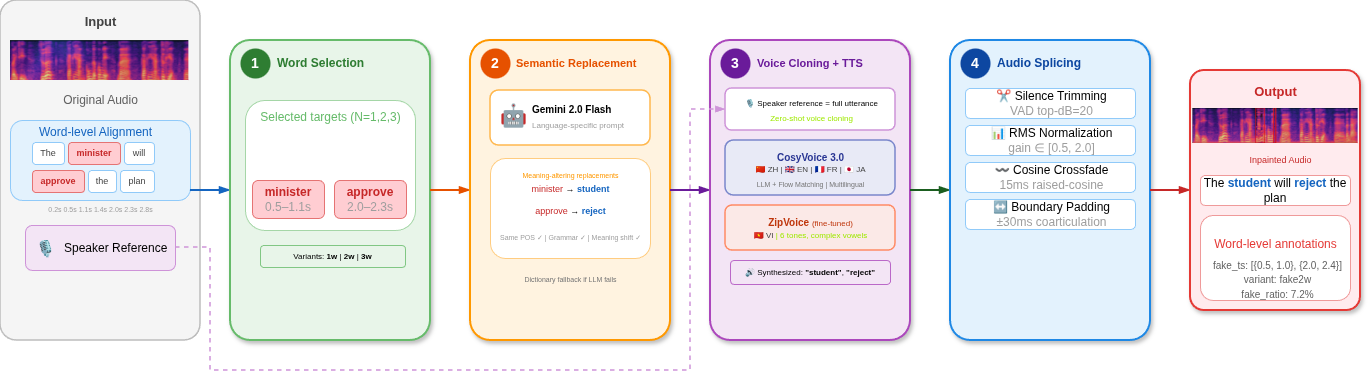}
    \caption{Overview of the MIST generation pipeline.
    Given a genuine utterance with word-level alignment from either
    Multilingual LibriSpeech (EN/FR/DE/IT/ES) or LEMAS-Dataset (VI),
    (1)~target words are selected based on duration and spacing constraints,
    (2)~semantically divergent replacements are generated via an LLM,
    (3)~replacement words are synthesized using speaker-conditioned voice
    cloning (CosyVoice~3 for EN/FR/DE/IT/ES or ZipVoice for VI),
    and (4)~synthesized segments are spliced into the original audio with
    crossfading and amplitude normalization.}
    \label{fig:pipeline}
\end{figure*}

The dataset generation pipeline, illustrated in Figure~\ref{fig:pipeline},
consists of four stages.
A critical design choice is that \textbf{voice cloning is performed
per-utterance before synthesis}: the TTS system first captures the speaker's
voice characteristics from the original recording, then generates the
replacement word in that voice.
This ensures maximum speaker consistency between the fake segment and its
surrounding context.

\textbf{Stage~1: Word selection.}
Given a source utterance with word-level timestamps provided by the
respective corpus's forced alignment, we select $N$ target words for
replacement ($N \in \{1,2,3\}$ depending on the variant).
Candidate words must satisfy three constraints:
(i)~minimum character length $\geq 3$ to avoid function words,
(ii)~minimum phonetic duration $\geq 150$\,ms to ensure sufficient acoustic
material for cloning, and
(iii)~minimum positional distance of 4 words between any two selected words
to avoid adjacent replacements that could merge into a single detectable
artifact.
A greedy selection algorithm with random shuffling is used, falling back to
relaxed constraints when the initial criteria are too restrictive.

\textbf{Stage~2: Semantic replacement via LLM.}
For each selected word, we generate a contextually appropriate replacement
using Gemini~2.0~Flash~\cite{team2024gemini} with a language-specific prompt.
The LLM is instructed to produce a single replacement word that
(i)~shares the same part of speech as the original,
(ii)~is grammatically correct within the sentence context,
(iii)~significantly alters the sentence's meaning, and
(iv)~is in the correct target language.
A dictionary-based fallback mechanism provides robustness when the LLM is
unavailable or returns malformed output.

\textbf{Stage~3: Speaker-conditioned voice cloning and synthesis.}
Each replacement word is synthesized using a zero-shot voice cloning TTS
model conditioned on the \emph{full original utterance} as a speaker
reference.
For English, French, German, Italian, and Spanish, we employ
\textbf{CosyVoice~3.0}~\cite{du2024cosyvoice,du2025cosyvoice3}, a
state-of-the-art multilingual zero-shot TTS system based on large language
models with flow matching.
For Vietnamese---which is not natively supported by CosyVoice---we use
\textbf{ZipVoice}, a TTS model fine-tuned specifically for Vietnamese speech
synthesis, to generate replacement words with appropriate tonal accuracy
and speaker characteristics.
This dual-model strategy ensures high synthesis quality across all six
languages.

\textbf{Stage~4: Audio splicing with artifact minimization.}
The synthesized replacement word is spliced into the original waveform at
the target word's temporal position.
To minimize audible artifacts at splice boundaries, we apply:
\begin{itemize}[leftmargin=*, itemsep=1pt]
    \item \textbf{Silence trimming}: energy-based VAD (top-dB\,=\,20) removes
          leading/trailing silence from the synthesized segment.
    \item \textbf{RMS normalization}: the amplitude of the synthesized segment
          is scaled to match the RMS energy of the original word
          (gain ratio clipped to $[0.5,\,2.0]$).
    \item \textbf{Cosine crossfading}: a 15\,ms raised-cosine fade is applied
          at both splice boundaries.
    \item \textbf{Padding}: a 30\,ms padding around the original word boundary
          accommodates coarticulation effects.
\end{itemize}

\subsection{Dataset Statistics}
\label{sec:dataset_stats}

Table~\ref{tab:dataset_comparison} compares MIST with existing partial
manipulation datasets.
MIST is the first dataset to provide (i)~multi-region word-level inpainting
labels with up to 3 tampered regions, (ii)~multilingual coverage across
6~languages, and (iii)~precise word-level temporal annotations for each
fake segment.

\begin{table*}[htbp]
\centering
\caption{Comparison of MIST with existing audio manipulation datasets.
``Max Regions'' indicates the maximum number of independently tampered
segments per utterance.
``Word-level'' indicates availability of word-level temporal annotations.}
\label{tab:dataset_comparison}
\small
\begin{tabular}{lcccc}
\toprule
\textbf{Dataset} & \textbf{\#Utt.} & \textbf{Max Regions}
                 & \textbf{Langs}  & \textbf{Word-level} \\
\midrule
ASVspoof~2019~\cite{wang2020asvspoof2019} & 121k & 0\,(full) & 1 & \texttimes \\
PartialSpoof~\cite{zhang2023partialspoof} & 108k & 1         & 1 & \texttimes \\
Half-Truth~\cite{yi2021halfthruth}        &  57k & 1         & 2 & \texttimes \\
LAV-DF~\cite{cai2022lavdf}               &  36k & 1         & 1 & \texttimes \\
LlamaPS~\cite{luong2024llamapartialspoof} & 130h & varies    & 1 & \texttimes \\
\midrule
\textbf{MIST (ours)} & \textbf{497k} & \textbf{3} & \textbf{6} & \textbf{\checkmark} \\
\bottomrule
\end{tabular}
\end{table*}

\begin{table*}[htbp]
\centering
\caption{Variant distribution in the MIST dataset (aggregated across all
languages). The fake ratio is defined as the total duration of tampered
segments divided by the utterance duration.}
\label{tab:variant_dist}
\small
\begin{tabular}{lccc}
\toprule
\textbf{Variant} & \textbf{\#Samples} & \textbf{Avg Duration (s)}
                 & \textbf{Avg Fake Ratio} \\
\midrule
1-word replacement  & 209,547 &  9.31 & 2.8\% \\
2-word replacement  & 213,428 &  9.18 & 5.7\% \\
3-word replacement  &  74,014 & 12.08 & 6.5\% \\
\midrule
Total fake          & 496,989 &  ---  & ---   \\
Total real (orig)   & 217,880 &  9.31 & 0\%   \\
\bottomrule
\end{tabular}
\end{table*}

\begin{table*}[htbp]
\centering
\caption{Per-language statistics of the MIST dataset.
Hours are computed from the total fake audio duration.
``Source Corpus'' and ``TTS Model'' indicate the data source and voice
cloning system used for each language.}
\label{tab:per_lang}
\small
\begin{tabular}{llrrrrll}
\toprule
\textbf{Lang} & \textbf{Name} & \textbf{\#Orig} & \textbf{\#Fake}
  & \textbf{Orig Hrs} & \textbf{Fake Hrs}
  & \textbf{Source Corpus} & \textbf{TTS Model} \\
\midrule
EN & English  &  44,937 &  99,362 & 126.0 & 279.3 & MLS (LibriSpeech) & CosyVoice3 \\
FR & French   &  30,523 &  75,172 &  83.0 & 207.6 & MLS (LibriSpeech) & CosyVoice3 \\
DE & German   &  36,812 &  83,204 &  97.0 & 218.4 & MLS (LibriSpeech) & CosyVoice3 \\
IT & Italian  &  31,748 &  73,892 &  79.0 & 192.1 & MLS (LibriSpeech) & CosyVoice3 \\
ES & Spanish  &  39,203 &  84,516 & 102.0 & 219.6 & MLS (LibriSpeech) & CosyVoice3 \\
VI & Vietnamese & 34,657 & 80,843 &  91.0 & 219.2 & LEMAS-Dataset     & ZipVoice (fine-tuned) \\
\midrule
\multicolumn{2}{l}{\textbf{Total}}
  & \textbf{217,880} & \textbf{496,989}
  & \textbf{578.0}   & \textbf{1{,}336.2} & --- & --- \\
\bottomrule
\end{tabular}
\end{table*}

\begin{figure}[htbp]
    \centering
    \includegraphics[width=\columnwidth]{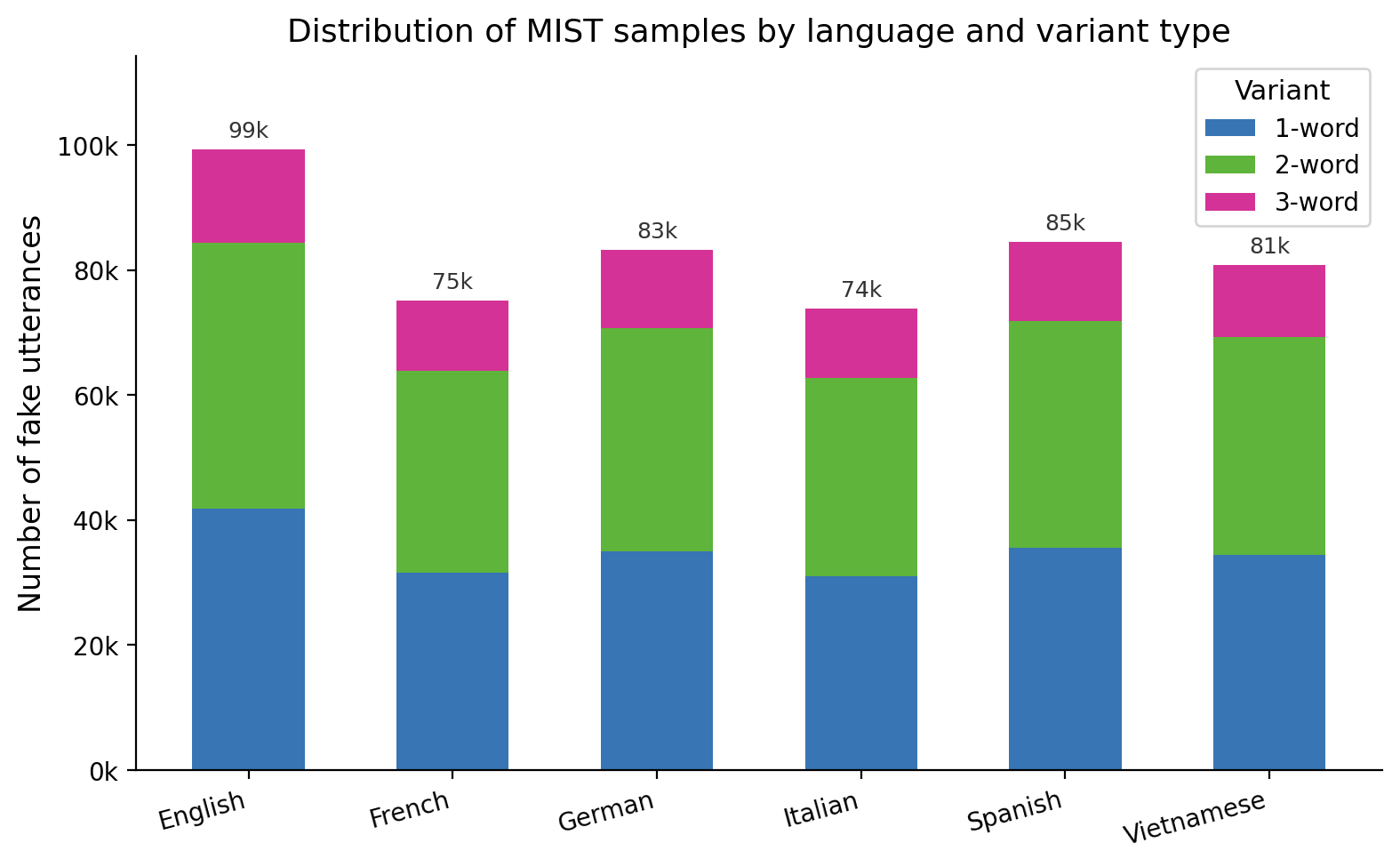}
    \caption{Distribution of MIST samples by language and variant type.
    Each language contributes approximately equal amounts of source data
    (${\sim}30$\,GB).
    The 3-word variant is only generated for utterances ${\geq}10$\,s,
    which explains its smaller share.}
    \label{fig:lang_dist}
\end{figure}

Table~\ref{tab:per_lang} presents the per-language breakdown.
Each language contributes approximately 30\,GB of source audio from its
respective corpus.
The number of fake variants varies across languages due to differences in
average utterance duration: languages with longer average utterances
(e.g., English, German) produce more 3-word variants, while languages with
shorter utterances (e.g., Italian, Vietnamese) produce proportionally more
1-word and 2-word variants, as visualised in Figure~\ref{fig:lang_dist}.

Table~\ref{tab:variant_dist} shows the variant distribution aggregated across
all languages.
The 1-word variant is the most abundant (present for all utterances), while
the 3-word variant is restricted to longer utterances ($\geq 10$\,s).
The average fake ratio increases predictably with the number of replaced
words, ranging from approximately 2.8\% for 1-word variants to 6.5\% for
3-word variants (Figure~\ref{fig:fake_ratio}).
This low fake ratio underscores the detection challenge: the vast majority
of each utterance remains genuine even in the hardest variant.

\begin{figure*}[htbp]
    \centering
    \includegraphics[width=\textwidth]{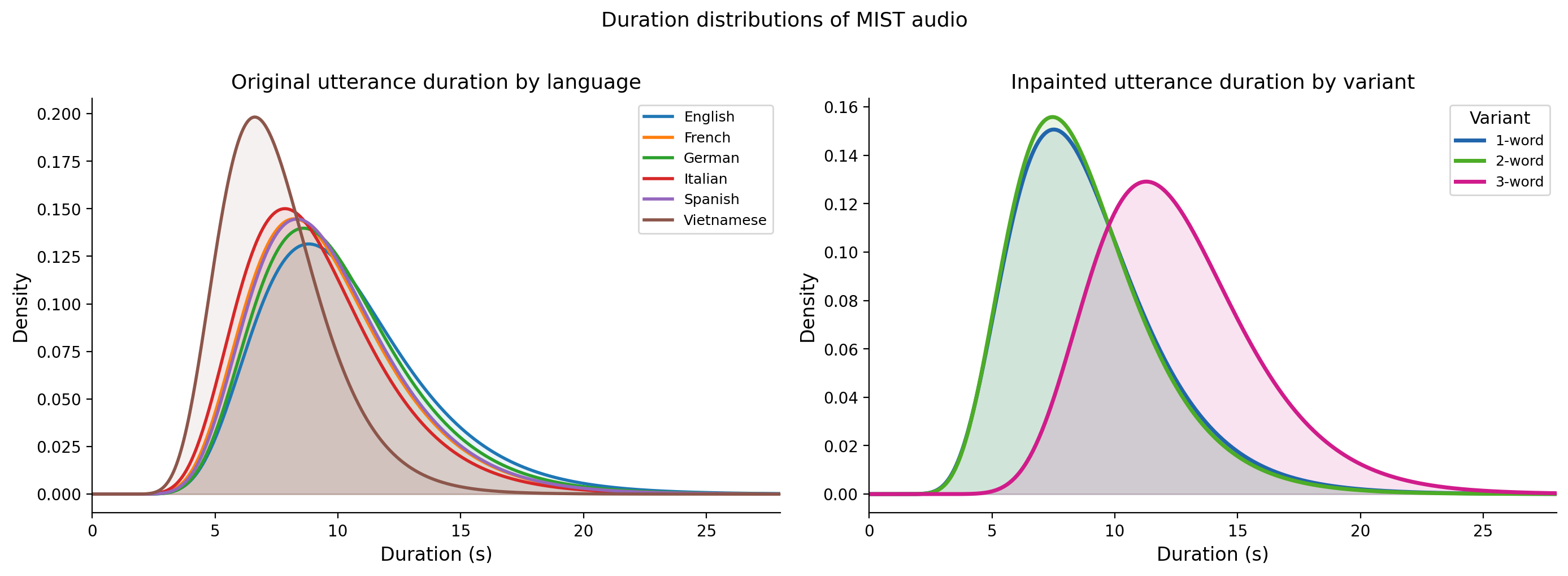}
    \caption{Duration distributions of MIST audio.
    \textbf{Left}: original utterance durations per language.\textbf{Right}: inpainted utterance durations by variant.}
    \label{fig:duration_dist}
\end{figure*}

\begin{figure*}[htbp]
    \centering
    \includegraphics[width=\textwidth]{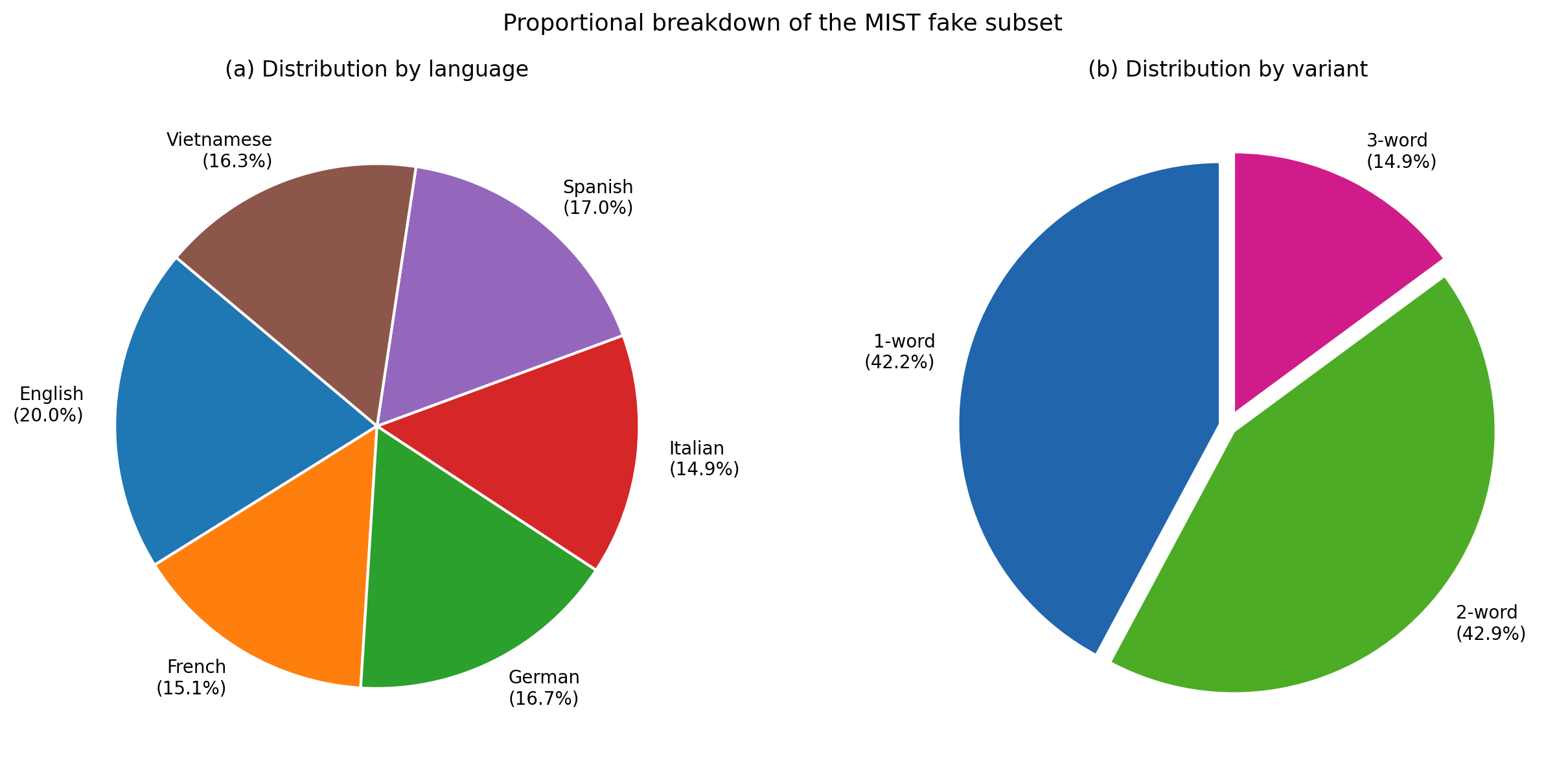}
    \caption{Proportional breakdown of the MIST fake subset.\textbf{(a)}~Distribution by language. \textbf{(b)}~Distribution by variant.}
    \label{fig:pie}
\end{figure*}

\begin{figure}[htbp]
    \centering
    \includegraphics[width=\columnwidth]{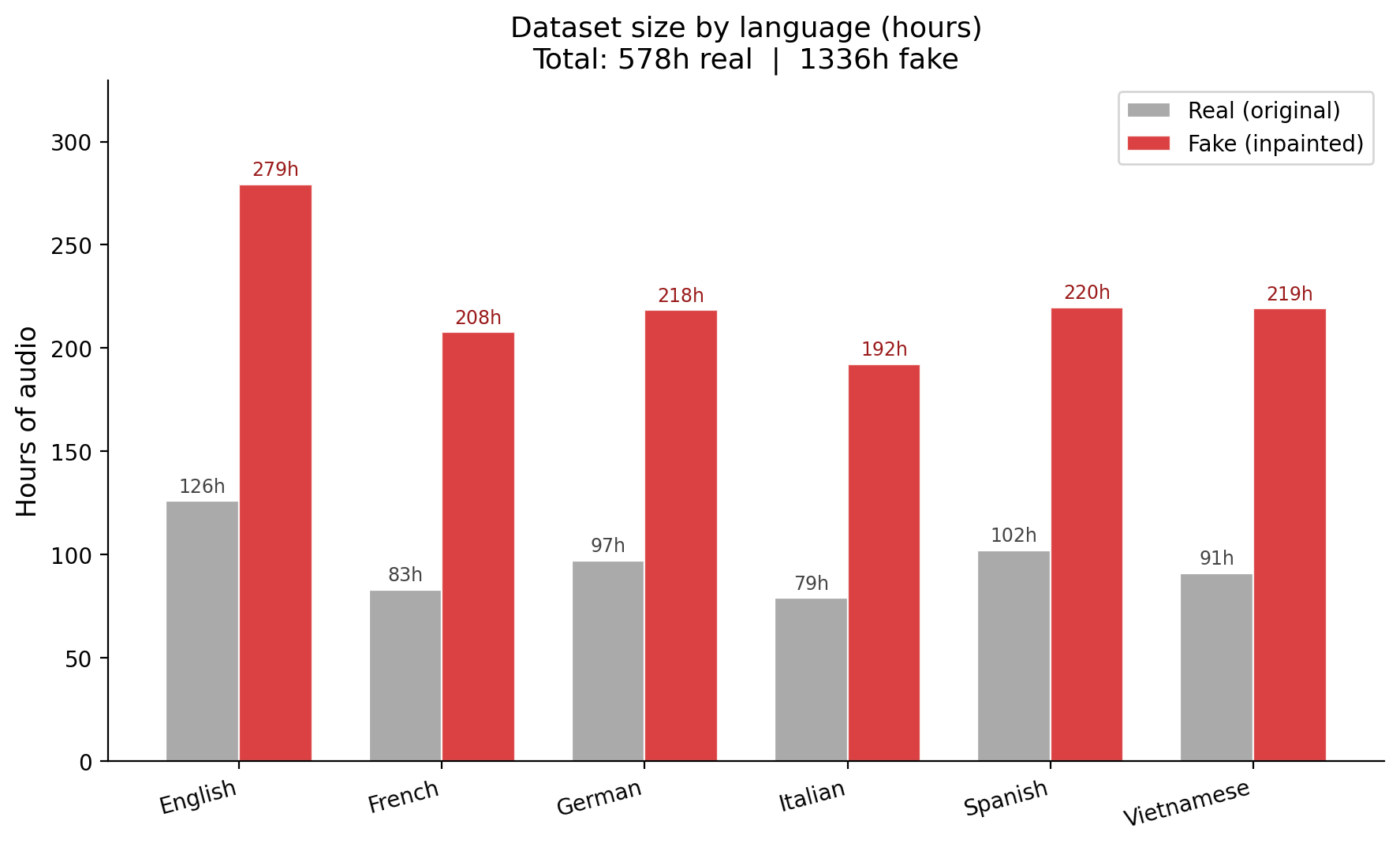}
    \caption{Dataset size by language (hours).Grey bars: original (real) audio. Red bars: inpainted (fake) audio.}
    \label{fig:hours_lang}
\end{figure}

\subsection{Multilingual Voice Cloning Strategy}
\label{sec:voice_cloning}

A key challenge in constructing a multilingual inpainting dataset is ensuring
high-quality, speaker-consistent synthesis across diverse languages.
We address this through a two-model strategy tailored to language coverage.

\textbf{CosyVoice~3.0 for EN, FR, DE, IT, ES.}
CosyVoice~3~\cite{du2024cosyvoice,du2025cosyvoice3} is a state-of-the-art
zero-shot TTS system that employs supervised semantic tokens derived from a
multilingual ASR model, combined with an LLM-based text-to-token generator
and a conditional flow-matching model for token-to-speech synthesis.
Its native multilingual support covers English, French, German, Italian, and
Spanish with high speaker similarity ($>0.85$ cosine similarity on speaker
embeddings) and content consistency.
For each replacement word, we provide the full original utterance as the
speaker reference and use instruction-following mode with a language-specific
prompt to ensure correct pronunciation and prosody.

\textbf{ZipVoice (fine-tuned) for VI.}
Vietnamese presents unique challenges for zero-shot TTS due to its six
lexical tones and complex vowel system.
Since CosyVoice does not natively support Vietnamese, we employ ZipVoice,
a TTS model fine-tuned on Vietnamese speech data, to generate replacement
words with appropriate tonal accuracy and speaker characteristics.

This dual-model approach ensures that each language receives synthesis from
a model specifically capable of handling its phonological characteristics,
resulting in consistently high-quality fake segments across all six languages.

\subsection{Quality Analysis}
\label{sec:quality}

We assess the quality of the generated dataset through both objective and
visual analyses.

\begin{figure*}[htbp]
    \centering
    \includegraphics[width=\textwidth]{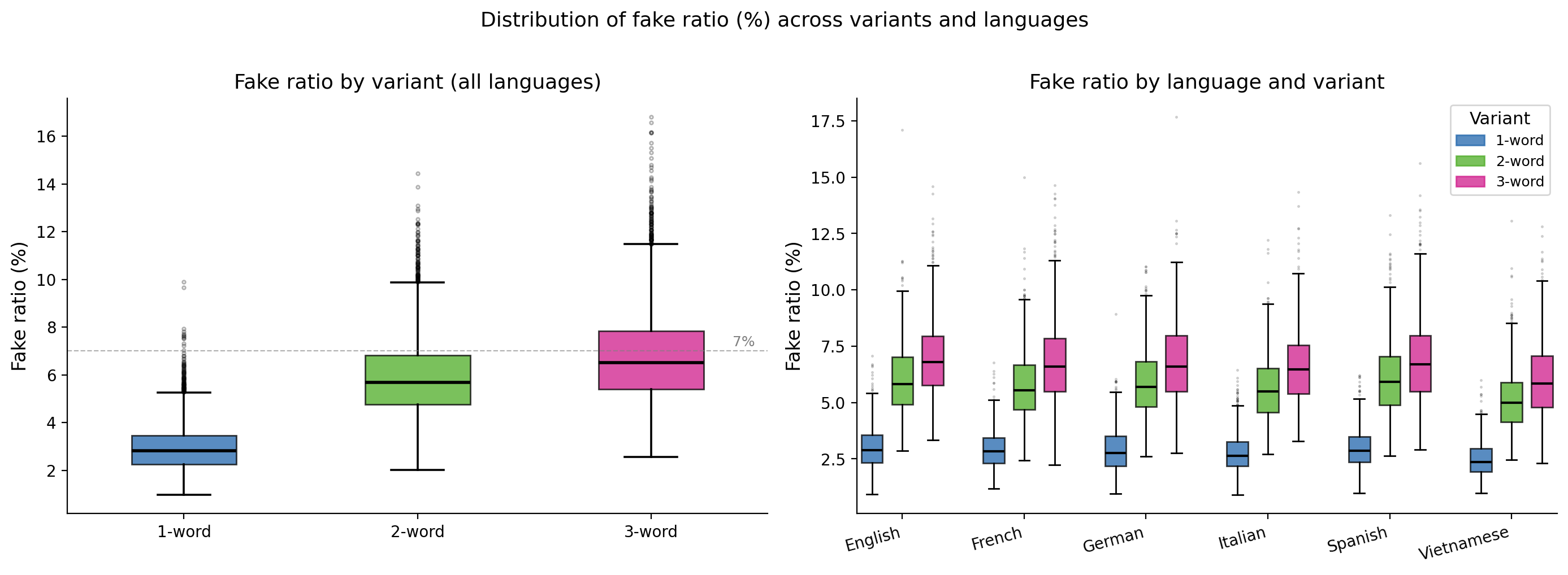}
    \caption{Distribution of fake ratio~(\%) by variant and language.}
    \label{fig:fake_ratio}
\end{figure*}

\textbf{Fake ratio analysis.}
Figure~\ref{fig:fake_ratio} shows the distribution of fake ratios across variants and languages.
The median fake ratio ranges from approximately 2.5\% for 1-word variants
to 6.5\% for 3-word variants, confirming that manipulated portions
constitute only a small fraction of each utterance.
Vietnamese tends to exhibit slightly lower fake ratios than the European
languages due to its shorter average word durations.

\begin{figure}[htbp]
    \centering
    \includegraphics[width=\columnwidth]{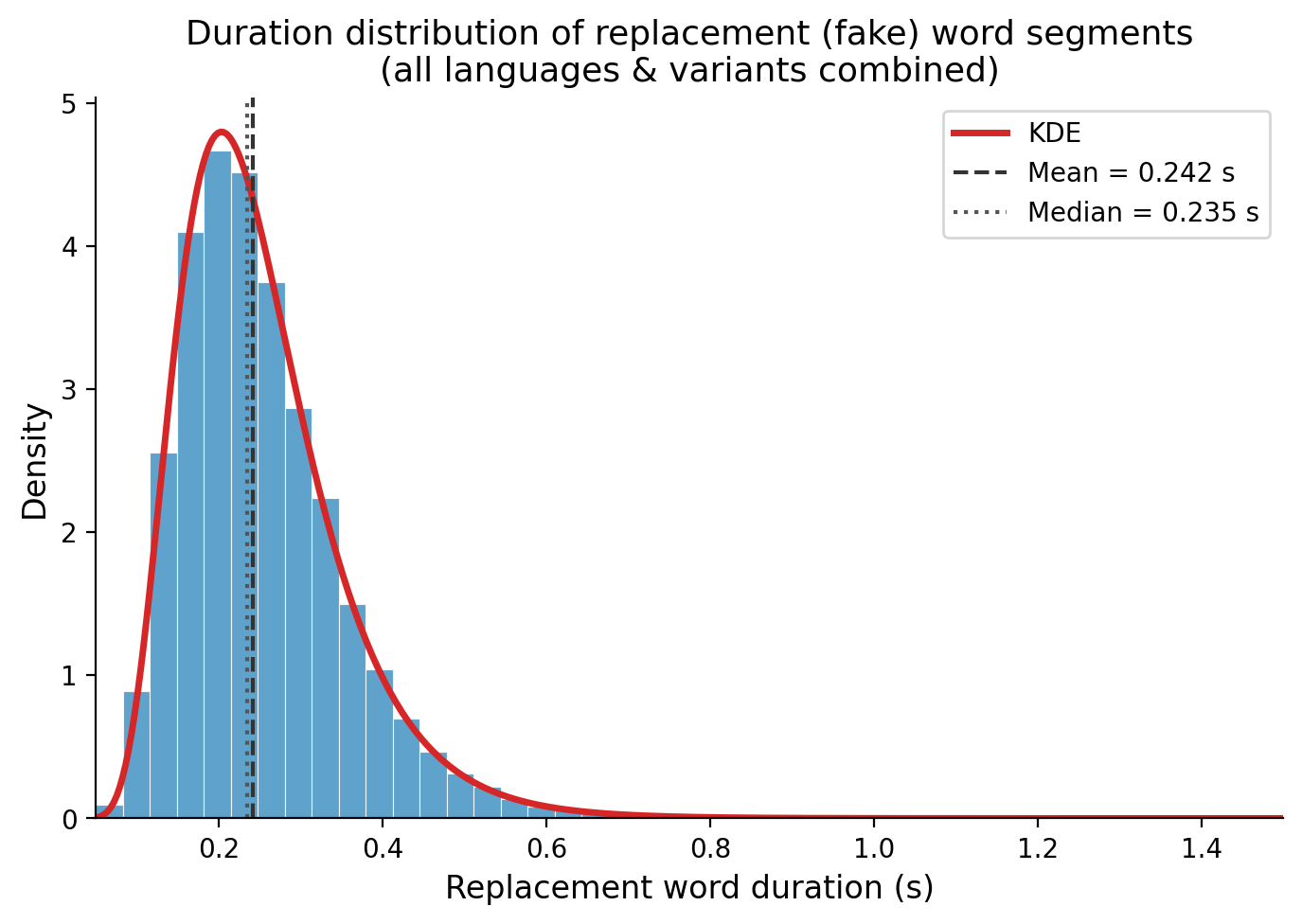}
    \caption{Duration distribution of individual replacement (fake) word
    segments across all languages and variants.}
    \label{fig:rep_duration}
\end{figure}

\textbf{Replacement word duration analysis.}
Figure~\ref{fig:rep_duration} shows the duration distribution of individual
replacement word segments.
The distribution is right-skewed, with a mean of 0.242\,s and median of
0.235\,s, consistent with natural spoken word durations across all six
languages.
The majority of segments fall between 0.1\,s and 0.5\,s, covering the
full range of short function words to longer content words.

\begin{figure*}[htbp]
    \centering
    \includegraphics[width=\textwidth]{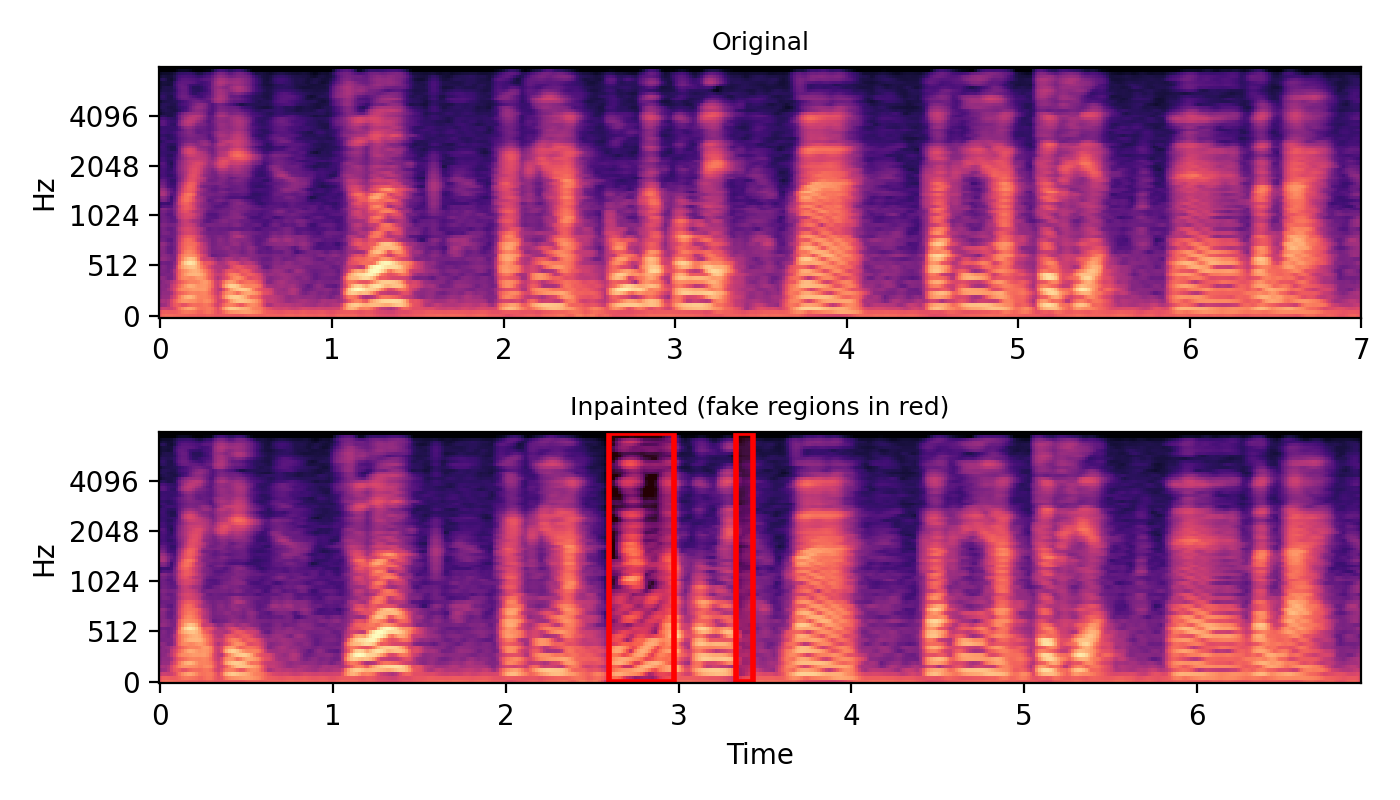}
    \caption{Mel-spectrogram comparison for an English utterance with
    2-word inpainting (fake2w variant).
    \textbf{Top}: original utterance.
    \textbf{Bottom}: inpainted utterance; red boxes mark the tampered regions.}
    \label{fig:spectrogram}
\end{figure*}

\textbf{Spectrogram analysis.}
Figure~\ref{fig:spectrogram} presents mel-spectrogram comparisons between
an original and its corresponding inpainted utterance.
The splice boundaries exhibit smooth energy transitions---attributable to
the 15\,ms cosine crossfading and RMS normalization steps---with no visible
discontinuities in the spectral envelope.
This visual seamlessness is indicative of the challenge posed to
spectrogram-based detectors.

\section{Iterative Segment Analysis}
\label{sec:method}

We propose \textbf{Iterative Segment Analysis (ISA)}, a backbone-agnostic framework that localizes an \emph{unknown} number of tampered regions in an audio signal through three successive stages of increasing granularity: coarse scanning, region proposal, and boundary refinement.

\subsection{Problem Formulation}
\label{sec:problem}

Let $\mathbf{x} \in \mathbb{R}^{L}$ denote a mono audio waveform of $L$ samples at sampling rate $r$ (Hz), corresponding to a total duration of $D = L / r$ seconds.
A tampered utterance contains $N \geq 1$ non-overlapping ground-truth fake segments
\begin{equation}
    \mathcal{S}^{*} = \bigl\{(s^{*}_n, e^{*}_n)\bigr\}_{n=1}^{N}, \quad 0 \leq s^{*}_n < e^{*}_n \leq D,
    \label{eq:gt_segments}
\end{equation}
where $s^{*}_n$ and $e^{*}_n$ are the start and end timestamps (in seconds) of the $n$-th manipulated region.
A genuine utterance has $N = 0$.
Crucially, the value of $N$ is \emph{unknown} at inference time and must be estimated jointly with the segment boundaries.

The localization task is to produce a set of $\hat{N}$ predicted segments
\begin{equation}
    \hat{\mathcal{S}} = \bigl\{(\hat{s}_m, \hat{e}_m)\bigr\}_{m=1}^{\hat{N}},
    \label{eq:pred_segments}
\end{equation}
that maximizes both the count accuracy ($\hat{N} \approx N$) and the temporal overlap with $\mathcal{S}^{*}$, as formalized by the SF1@$\tau$ metric introduced in Section~\ref{sec:metric}.

\subsection{Method Overview}
\label{sec:isa_overview}

\begin{figure*}[htbp]
    \centering
    \includegraphics[width=\textwidth]{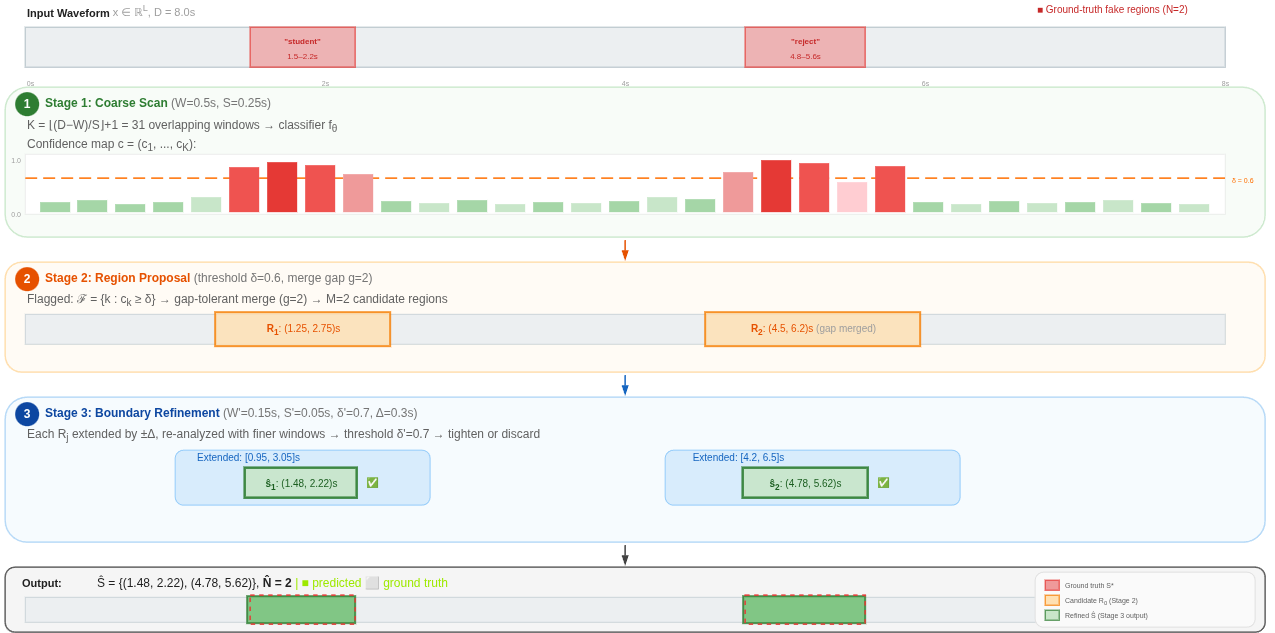}
    \caption{Iterative Segment Analysis (ISA) pipeline illustrated on a 2-word inpainted utterance.
    \textbf{Stage~1:} A sliding window ($W{=}0.5$\,s, $S{=}0.25$\,s) produces a coarse confidence map; windows exceeding $\delta{=}0.6$ are flagged (red).
    \textbf{Stage~2:} Flagged windows are merged with gap tolerance $g{=}2$, yielding candidate regions (orange boxes).
    \textbf{Stage~3:} Each candidate is re-analyzed with finer windows ($W'{=}0.15$\,s, $S'{=}0.05$\,s) and threshold $\delta'{=}0.7$; boundaries are tightened to the refined extent (green boxes). False positive candidates are discarded.}
    \label{fig:isa_method}
\end{figure*}

ISA decomposes the localization problem into three stages (Figure~\ref{fig:isa_method}):

\begin{enumerate}[leftmargin=*, itemsep=2pt, label=\textbf{Stage \arabic*:}]
    \item \textbf{Coarse Scan} --- A sliding window with large window size sweeps across the waveform; a binary classifier scores each window, producing a frame-level \emph{confidence map}.
    \item \textbf{Region Proposal} --- The confidence map is thresholded and clustered into contiguous candidate regions via gap-tolerant merging.
    \item \textbf{Boundary Refinement} --- Each candidate region is re-analyzed at finer temporal resolution to tighten its boundaries and filter false positives.
\end{enumerate}

\noindent
The key insight is that a single forward pass of a deepfake classifier over the full utterance cannot resolve individual tampered words (which may last only 0.2--0.8\,s).
By iterating from coarse to fine, ISA first identifies \emph{where} to look, then precisely \emph{delineates} each region, achieving high recall without excessive computational cost.

\subsection{Stage 1: Coarse Scan}
\label{sec:coarse}

Let $f_\theta : \mathbb{R}^{W \cdot r} \to [0, 1]$ denote a binary deepfake classifier parameterized by $\theta$, which accepts an audio segment of duration $W$ seconds (i.e., $W \cdot r$ samples) and outputs a scalar confidence $c \in [0, 1]$ representing the estimated probability that the segment contains manipulated content.

We partition $\mathbf{x}$ into $K$ overlapping windows using window size $W$ and stride $S$ ($S < W$ to ensure overlap):
\begin{equation}
    K = \left\lfloor \frac{D - W}{S} \right\rfloor + 1.
    \label{eq:num_windows}
\end{equation}
The $k$-th window ($k = 1, \ldots, K$) spans the time interval
\begin{equation}
    \bigl[t_k,\; t_k + W\bigr], \quad t_k = (k - 1) \cdot S,
    \label{eq:window_time}
\end{equation}
where $t_k$ is the left edge of window $k$.
Each window is independently classified:
\begin{equation}
    c_k = f_\theta\!\bigl(\mathbf{x}[t_k \cdot r : (t_k + W) \cdot r]\bigr),
    \label{eq:confidence}
\end{equation}
yielding the \emph{confidence map} $\mathbf{c} = (c_1, c_2, \ldots, c_K) \in [0,1]^K$.

\textbf{Intuition.}
Windows that overlap entirely with a genuine region will receive low confidence ($c_k \approx 0$), while windows containing even partial fake content tend to produce elevated scores.
The overlap between adjacent windows (ratio $1 - S/W$) provides redundancy that smooths sporadic misclassifications.

\subsection{Stage 2: Region Proposal and Merging}
\label{sec:proposal}

We convert the confidence map into discrete candidate regions through thresholding and merging.

\textbf{Step 2a: Thresholding.}
A window is flagged as \emph{suspicious} if its confidence exceeds a detection threshold~$\delta$:
\begin{equation}
    \mathcal{F} = \bigl\{k : c_k \geq \delta\bigr\}.
    \label{eq:flagged}
\end{equation}

\textbf{Step 2b: Gap-tolerant merging.}
Consecutive flagged windows naturally form contiguous runs.
However, a single missed window between two true positives would incorrectly split one tampered region into two.
To address this, we introduce a \emph{merge gap tolerance}~$g$: if two flagged runs are separated by at most $g$ unflagged windows, they are merged into a single candidate region.

Formally, we sort $\mathcal{F} = \{k_1, k_2, \ldots\}$ in ascending order and group elements into clusters $\mathcal{G}_1, \mathcal{G}_2, \ldots$ such that consecutive elements within a cluster satisfy $k_{i+1} - k_i \leq g + 1$.
Each cluster $\mathcal{G}_j$ is mapped to a candidate region by converting window indices back to timestamps:
\begin{equation}
    \mathcal{R}_0 = \left\{\bigl(t_{\min(\mathcal{G}_j)},\; t_{\max(\mathcal{G}_j)} + W\bigr)\right\}_{j=1}^{M},
    \label{eq:candidates}
\end{equation}
where $M = |\{\mathcal{G}_j\}|$ is the number of candidate regions, and $t_k$ is defined in Eq.~\eqref{eq:window_time}.

\textbf{Early termination.}
If $\mathcal{F} = \emptyset$ (no window exceeds $\delta$), the utterance is classified as entirely genuine: $\hat{\mathcal{S}} = \emptyset$, $\hat{N} = 0$.

\subsection{Stage 3: Boundary Refinement}
\label{sec:refine}

The coarse scan localizes tampered regions to within approximately $\pm W/2$ seconds.
To achieve word-level precision, we re-analyze each candidate at finer granularity.

For each candidate region $(s_j, e_j) \in \mathcal{R}_0$, we define an \emph{extended analysis interval}:
\begin{equation}
    \bigl[\tilde{s}_j,\; \tilde{e}_j\bigr] = \bigl[\max(0,\; s_j - \Delta),\; \min(D,\; e_j + \Delta)\bigr],
    \label{eq:extended}
\end{equation}
where $\Delta$ is the \emph{boundary extension margin} (in seconds) that ensures the true boundaries lie within the analysis window.

Within this interval, we apply the same classifier $f_\theta$ with a finer window size $W'$ and stride $S'$ ($W' < W$, $S' < S$), producing a refined confidence map $\mathbf{c}' = (c'_1, \ldots, c'_{K'_j})$ over $K'_j$ sub-windows.
Specifically:
\begin{equation}
    K'_j = \left\lfloor \frac{(\tilde{e}_j - \tilde{s}_j) - W'}{S'} \right\rfloor + 1.
    \label{eq:fine_windows}
\end{equation}

\textbf{Step 3a: Refined thresholding.}
We apply a (typically stricter) refinement threshold $\delta'$ ($\delta' \geq \delta$) to the fine-grained confidence map:
\begin{equation}
    \mathcal{F}'_j = \bigl\{k : c'_k \geq \delta'\bigr\}.
    \label{eq:fine_flagged}
\end{equation}

\textbf{Step 3b: False positive suppression.}
If $\mathcal{F}'_j = \emptyset$---i.e., no fine-grained window exceeds $\delta'$---the candidate region $(s_j, e_j)$ is discarded as a false positive from the coarse stage.

\textbf{Step 3c: Boundary tightening.}
For surviving candidates, the refined boundaries are set to the temporal extent of the first and last flagged fine-grained windows:
\begin{equation}
    (\hat{s}_j, \hat{e}_j) = \bigl(\tilde{s}_j + (\min \mathcal{F}'_j - 1) \cdot S',\;\; \tilde{s}_j + (\max \mathcal{F}'_j - 1) \cdot S' + W'\bigr).
    \label{eq:refined_bounds}
\end{equation}

The final output is the refined segment set $\hat{\mathcal{S}} = \{(\hat{s}_j, \hat{e}_j) : \mathcal{F}'_j \neq \emptyset\}$, with $\hat{N} = |\hat{\mathcal{S}}|$.

\subsection{Backbone Classifier}
\label{sec:backbone}

ISA treats $f_\theta$ as a black-box scoring function, making it compatible with any audio deepfake detector that accepts a fixed-length waveform segment and outputs a spoofing probability.
In our experiments, we evaluate three architectures spanning different feature extraction paradigms:

\begin{itemize}[leftmargin=*, itemsep=3pt]
    \item \textbf{Wav2Vec2-AASIST.}
    Self-supervised Wav2Vec 2.0~\cite{baevski2020wav2vec2} features are extracted from the input waveform and passed to the AASIST~\cite{jung2022aasist} graph attention network, which models spectro-temporal dependencies via heterogeneous attention.
    This combination leverages large-scale pre-trained representations with a purpose-built anti-spoofing classifier.

    \item \textbf{WavLM-AASIST.}
    WavLM~\cite{chen2022wavlm}, a self-supervised model pre-trained with both masked speech prediction and speaker-aware objectives, replaces Wav2Vec 2.0 as the feature extractor.
    The richer speaker-discriminative representations may benefit detection of speaker-cloned content.

    \item \textbf{Wav2Vec2-Linear.}
    Wav2Vec 2.0 features are classified by a single linear layer~\cite{tak2022wav2vec_spoofing}.
    This minimal architecture serves as a lower-bound baseline, isolating the contribution of the ISA framework itself from the backbone's capacity.
\end{itemize}

\noindent
All backbones are trained on \emph{utterance-level} binary labels (real vs.\ fake) using the standard cross-entropy loss.
No frame-level or segment-level annotations are used during training---ISA enables segment-level localization purely at inference time by querying the utterance-level classifier on sub-utterance windows.
This is a significant practical advantage, as segment-level labels are costly to obtain at scale.

\subsection{Implementation Details}
\label{sec:impl}

Table~\ref{tab:hyperparams} summarizes the ISA hyperparameters, which were selected via grid search on a held-out validation set from the MIST dataset (Section~\ref{sec:dataset}).

\begin{table*}[htbp]
\centering
\caption{ISA hyperparameters. The coarse stage (Stage~1) uses larger windows for efficient scanning; the refinement stage (Stage~3) uses smaller windows for precise boundary delineation.}
\label{tab:hyperparams}
\small
\begin{tabular}{llccl}
\toprule
\textbf{Symbol} & \textbf{Parameter} & \textbf{Coarse (Stage 1)} & \textbf{Fine (Stage 3)} & \textbf{Description} \\
\midrule
$W$ / $W'$          & Window size       & 0.5\,s  & 0.15\,s & Duration of each analysis window \\
$S$ / $S'$          & Stride            & 0.25\,s & 0.05\,s & Step between consecutive windows \\
$\delta$ / $\delta'$ & Threshold         & 0.6     & 0.7     & Minimum confidence to flag a window \\
$g$                  & Merge gap         & 2       & ---     & Max unflagged windows between merged runs \\
$\Delta$             & Extension margin  & ---     & 0.3\,s  & Padding around candidate for re-analysis \\
\bottomrule
\end{tabular}
\end{table*}

\textbf{Window sizing rationale.}
The coarse window $W = 0.5$\,s is chosen to be comparable to the average replacement word duration in the MIST dataset (0.3--0.6\,s), ensuring that at least one coarse window is dominated by fake content for each tampered word.
The fine window $W' = 0.15$\,s provides sub-word resolution, enabling boundary precision of approximately $\pm S' = \pm 0.05$\,s.

\textbf{Threshold selection.}
The coarse threshold $\delta = 0.6$ is set conservatively (below the typical decision boundary of 0.5 used in utterance-level classification) to favor recall over precision at the proposal stage.
The refinement threshold $\delta' = 0.7$ is stricter, suppressing false positives that survived the coarse stage.

\textbf{Merge gap rationale.}
A gap tolerance of $g = 2$ windows corresponds to a temporal gap of $g \cdot S = 0.5$\,s.
This prevents splitting a single tampered word into multiple fragments due to isolated low-confidence windows, while remaining small enough to avoid merging two distinct tampered regions that are separated by at least 4 words (typically $> 1.5$\,s apart due to the word-spacing constraint in Section~\ref{sec:pipeline}).

\textbf{Computational cost.}
ISA's computational overhead beyond the backbone classifier is negligible: Stage~2 and Stage~3 involve only thresholding, sorting, and index arithmetic.
The dominant cost is the $K + \sum_j K'_j$ forward passes of $f_\theta$.
For a typical 10\,s utterance with the default hyperparameters, the coarse stage requires $K = 39$ inferences and each refinement region requires $K'_j \approx 20$ inferences, yielding fewer than 100 total classifier calls per utterance.
With batched inference on a single GPU, the total ISA pipeline processes one utterance in under 0.3\,s.

\textbf{Training details.}
Each backbone $f_\theta$ is trained for 20 epochs on the MIST training set using the AdamW optimizer with an initial learning rate of $10^{-4}$ and cosine annealing.
The input is a randomly cropped $W$-second segment: for fake utterances, a segment overlapping a tampered region is sampled with probability 0.5 (balanced sampling).
Data augmentation includes additive Gaussian noise ($\text{SNR} \in [15, 30]$\,dB) and random gain perturbation ($\pm 3$\,dB).
All audio is resampled to 16\,kHz mono.

\section{Evaluation Metric: SF1@$\tau$}
\label{sec:metric}

Existing audio deepfake evaluation protocols rely on utterance-level or frame-level metrics, neither of which adequately captures the multi-region localization task addressed in this work.
We propose \textbf{SF1@$\tau$}, a segment-level F1 score based on temporal Intersection-over-Union (IoU) matching, directly inspired by the mean Average Precision (mAP@$\tau$) metric used in object detection~\cite{cai2022lavdf}.

\subsection{Limitations of Existing Metrics}
\label{sec:metric_limits}

We identify three categories of existing metrics and their limitations for the multi-region localization setting:

\textbf{Utterance-level metrics} (accuracy, Equal Error Rate).
These classify entire utterances as real or fake.
They provide no information about \emph{where} or \emph{how many} regions are tampered, and assign the same score to a detector that correctly localizes two fake words as to one that blindly labels the entire utterance as fake.

\textbf{Frame-level metrics} (per-frame AUC, frame accuracy).
These evaluate each time frame independently, treating the prediction as a binary segmentation mask.
While they capture some spatial information, they suffer from two critical shortcomings:
(i)~they do not penalize \emph{fragmentation}---a single tampered region predicted as multiple disjoint fragments receives the same score as a single correct prediction, and
(ii)~they are dominated by the majority class (genuine frames typically constitute $>90\%$ of each utterance), inflating scores without reflecting true localization quality.

\textbf{Boundary-based metrics} (onset/offset error).
These measure the temporal distance between predicted and true boundaries but require a pre-defined one-to-one correspondence between predictions and ground truths.
They are ill-suited when the number of predicted segments $\hat{N}$ differs from the true count $N$, which is the common case in practice.

These limitations motivate a metric that jointly evaluates three aspects: (i)~segment \emph{count} estimation, (ii)~segment \emph{position} accuracy, and (iii)~segment \emph{boundary} precision.

\subsection{Temporal Intersection-over-Union}
\label{sec:iou}

We first define the temporal overlap measure between a predicted segment and a ground-truth segment.
Recall from Section~\ref{sec:problem} that the ground-truth segments are $\mathcal{S}^{*} = \{(s^{*}_n, e^{*}_n)\}_{n=1}^{N}$ and the predicted segments are $\hat{\mathcal{S}} = \{(\hat{s}_m, \hat{e}_m)\}_{m=1}^{\hat{N}}$.

For a predicted segment $\hat{\sigma}_m = (\hat{s}_m, \hat{e}_m)$ and a ground-truth segment $\sigma^{*}_n = (s^{*}_n, e^{*}_n)$, both representing time intervals on $[0, D]$, the \emph{temporal IoU} is defined as:
\begin{equation}
    \operatorname{IoU}(\hat{\sigma}_m, \sigma^{*}_n) = \frac{|\hat{\sigma}_m \cap \sigma^{*}_n|}{|\hat{\sigma}_m \cup \sigma^{*}_n|},
    \label{eq:iou}
\end{equation}
where $|\cdot|$ denotes the duration (in seconds) of a time interval, the intersection is
\begin{equation}
    |\hat{\sigma}_m \cap \sigma^{*}_n| = \max\!\bigl(0,\; \min(\hat{e}_m, e^{*}_n) - \max(\hat{s}_m, s^{*}_n)\bigr),
    \label{eq:intersection}
\end{equation}
and the union follows from the inclusion-exclusion principle:
\begin{equation}
    |\hat{\sigma}_m \cup \sigma^{*}_n| = |\hat{\sigma}_m| + |\sigma^{*}_n| - |\hat{\sigma}_m \cap \sigma^{*}_n|.
    \label{eq:union}
\end{equation}

The IoU takes values in $[0, 1]$: a value of 0 indicates no temporal overlap, while 1 indicates perfect alignment.

\subsection{Greedy Bipartite Matching}
\label{sec:matching}

Given a threshold $\tau \in (0, 1]$, we define a matching between $\hat{\mathcal{S}}$ and $\mathcal{S}^{*}$ to determine which predictions correspond to true tampered regions.

\textbf{Matching procedure.}
We construct the $\hat{N} \times N$ IoU matrix $\mathbf{A}$ with entries $A_{mn} = \operatorname{IoU}(\hat{\sigma}_m, \sigma^{*}_n)$.
A greedy one-to-one matching is performed as follows:

\begin{enumerate}[leftmargin=*, itemsep=2pt, label=(\roman*)]
    \item Identify the maximum entry $A_{m^*n^*} = \max_{m,n} A_{mn}$ among all unmatched pairs.
    \item If $A_{m^*n^*} \geq \tau$, match $\hat{\sigma}_{m^*}$ to $\sigma^{*}_{n^*}$; mark both as matched.
    \item Repeat steps (i)--(ii) until no unmatched pair satisfies $A_{mn} \geq \tau$.
\end{enumerate}

\noindent
Each ground-truth segment is matched to \emph{at most one} predicted segment and vice versa, ensuring that neither over-segmentation (multiple predictions covering one ground truth) nor under-segmentation (one prediction covering multiple ground truths) is rewarded.

Let $\mathcal{M} \subseteq \{1, \ldots, \hat{N}\} \times \{1, \ldots, N\}$ denote the resulting set of matched pairs.

\subsection{SF1@$\tau$ Computation}
\label{sec:sf1_compute}

From the matching $\mathcal{M}$, we compute segment-level precision, recall, and F1 for a single utterance:

\begin{align}
    \mathrm{TP} &= |\mathcal{M}|, \label{eq:tp} \\
    \mathrm{FP} &= \hat{N} - \mathrm{TP}, \label{eq:fp} \\
    \mathrm{FN} &= N - \mathrm{TP}, \label{eq:fn}
\end{align}
where $\mathrm{TP}$ counts correctly localized predictions, $\mathrm{FP}$ counts spurious predictions (false alarms or mislocalized segments), and $\mathrm{FN}$ counts missed ground-truth regions.

The segment-level precision ($\mathrm{SP}$), recall ($\mathrm{SR}$), and F1 for a single utterance are:
\begin{align}
    \mathrm{SP}@\tau &= \frac{\mathrm{TP}}{\mathrm{TP} + \mathrm{FP}} = \frac{|\mathcal{M}|}{\hat{N}}, \label{eq:sprec} \\[4pt]
    \mathrm{SR}@\tau &= \frac{\mathrm{TP}}{\mathrm{TP} + \mathrm{FN}} = \frac{|\mathcal{M}|}{N}, \label{eq:srec} \\[4pt]
    \mathrm{SF1}@\tau &= \frac{2 \cdot \mathrm{SP}@\tau \cdot \mathrm{SR}@\tau}{\mathrm{SP}@\tau + \mathrm{SR}@\tau}. \label{eq:sf1}
\end{align}

\textbf{Edge cases.}
For a genuine utterance ($N = 0$): if $\hat{N} = 0$, the utterance is a true negative and excluded from the F1 average (contributing only to CA below); if $\hat{N} > 0$, all predictions are false positives, and SF1@$\tau = 0$.
For a fake utterance ($N \geq 1$): if $\hat{N} = 0$, then $\mathrm{TP} = 0$ and SF1@$\tau = 0$.

\textbf{Aggregation.}
The dataset-level SF1@$\tau$ is the \emph{macro-average} over all utterances containing at least one tampered region:
\begin{equation}
    \overline{\mathrm{SF1}}@\tau = \frac{1}{|\mathcal{D}_{\text{fake}}|} \sum_{u \in \mathcal{D}_{\text{fake}}} \mathrm{SF1}@\tau_u,
    \label{eq:sf1_macro}
\end{equation}
where $\mathcal{D}_{\text{fake}} = \{u \in \mathcal{D} : N_u \geq 1\}$ is the set of tampered utterances in the evaluation set $\mathcal{D}$.

\textbf{Primary and lenient thresholds.}
We report two threshold settings:
\begin{itemize}[leftmargin=*, itemsep=2pt]
    \item \textbf{SF1@0.5} (primary): a predicted segment must overlap at least 50\% with a ground-truth segment (IoU $\geq 0.5$) to count as a true positive. This is a standard strictness level analogous to mAP@0.5 in object detection.
    \item \textbf{SF1@0.3} (lenient): a 30\% IoU threshold that credits coarser but directionally correct localizations, useful for evaluating methods with less precise boundary estimation.
\end{itemize}

\subsection{Complementary Metric: Count Accuracy}
\label{sec:ca}

SF1@$\tau$ conflates two sources of error: incorrect segment \emph{count} and inaccurate segment \emph{boundaries}.
To disentangle these, we introduce \textbf{Count Accuracy (CA)}, which evaluates only the count estimation aspect:
\begin{equation}
    \mathrm{CA} = \frac{1}{|\mathcal{D}|} \sum_{u \in \mathcal{D}} \mathbbm{1}\!\left[\hat{N}_u = N_u\right],
    \label{eq:ca}
\end{equation}
where $\mathbbm{1}[\cdot]$ is the indicator function and the sum runs over \emph{all} utterances in the evaluation set (including genuine ones with $N_u = 0$).

CA measures how often a system correctly estimates the number of tampered regions, regardless of their temporal accuracy.
A system with high CA but low SF1@$\tau$ identifies the right number of fake segments but localizes them poorly; conversely, high SF1@$\tau$ with low CA is impossible by construction (since miscounting necessarily generates FP or FN).

\subsection{Relation to Object Detection Metrics}
\label{sec:metric_compare}

SF1@$\tau$ is a specialization of the mAP framework from visual object detection~\cite{cai2022lavdf} to the one-dimensional temporal domain.
Table~\ref{tab:metric_analogy} summarizes the analogy.

\begin{table*}[htbp]
\centering
\caption{Analogy between SF1@$\tau$ (proposed) and mAP@$\tau$ from object detection. SF1@$\tau$ adapts the spatial IoU matching paradigm to one-dimensional temporal segments.}
\label{tab:metric_analogy}
\small
\begin{tabular}{lll}
\toprule
\textbf{Concept} & \textbf{Object Detection (2D)} & \textbf{Audio Inpainting (1D)} \\
\midrule
Prediction unit    & Bounding box $(x, y, w, h)$             & Time interval $(\hat{s}, \hat{e})$ \\
Ground truth       & Annotated object box                     & Annotated tampered segment $(s^{*}, e^{*})$ \\
Overlap measure    & Spatial IoU (area)                       & Temporal IoU (duration), Eq.~\eqref{eq:iou} \\
Matching           & Greedy or Hungarian, per class           & Greedy bipartite, single class (fake) \\
Threshold          & $\tau \in \{0.5, 0.75, 0.5{:}0.95\}$   & $\tau \in \{0.3, 0.5\}$ \\
Aggregation        & AP per class $\to$ mAP                   & F1 per utterance $\to$ macro-average \\
Complementary      & Object count error                       & Count Accuracy (CA), Eq.~\eqref{eq:ca} \\
\bottomrule
\end{tabular}
\end{table*}

Two key differences from standard mAP are worth noting.
First, we use F1 rather than AP (area under the precision-recall curve) because the deepfake detector outputs a binary decision per segment rather than a continuous ranking.
Second, our task involves a single class (``fake''), eliminating the need for per-class averaging.

\textbf{Why not frame-level F1?}
One might consider computing F1 at the frame level (each 10\,ms frame labeled real/fake) as in prior work~\cite{zhang2023partialspoof}.
However, frame-level F1 does not penalize fragmentation: a single tampered word predicted as 5 tiny fragments and one correct prediction yield identical frame-level TP counts.
SF1@$\tau$ explicitly penalizes this via the IoU threshold, which requires each prediction to substantially overlap a \emph{single} contiguous ground-truth region.

\subsection{Illustrative Example}
\label{sec:metric_example}

\begin{figure*}[htbp]
    \centering
    \includegraphics[width=\textwidth]{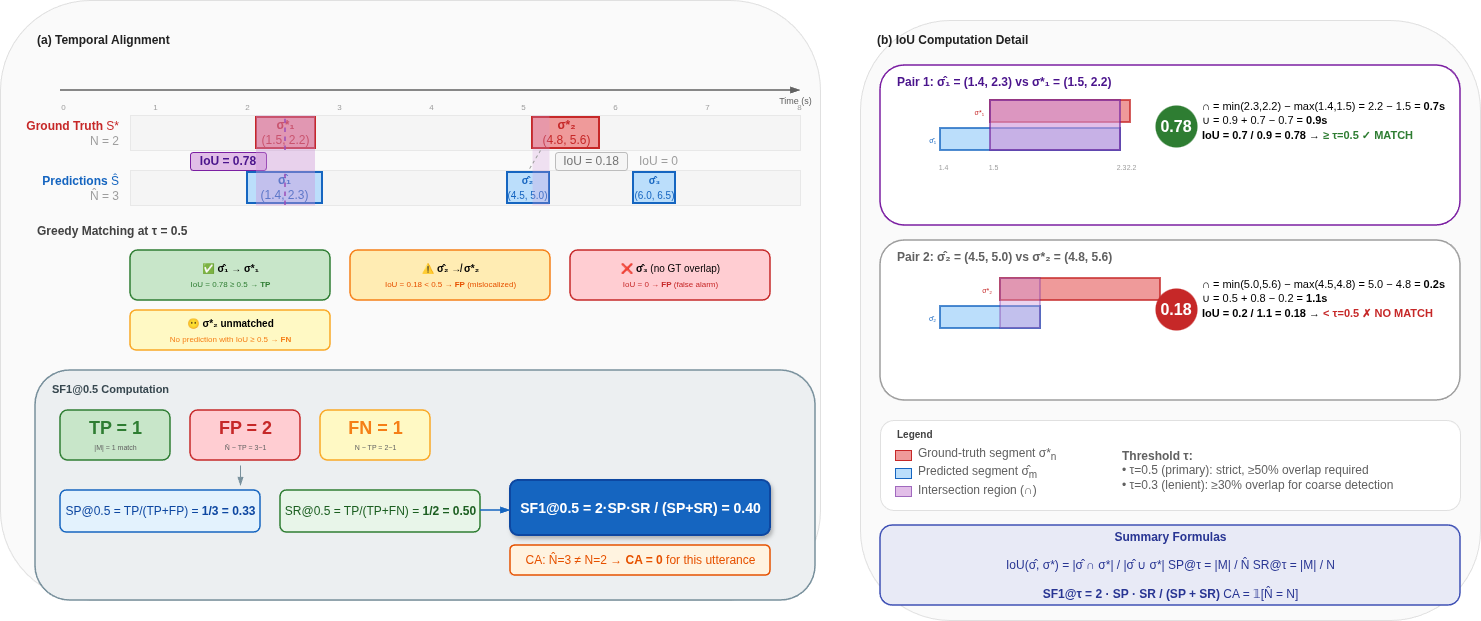}
    \caption{Illustrative example of SF1@$\tau$ computation.
    An utterance with $N{=}2$ ground-truth segments (red) receives $\hat{N}{=}3$ predictions (blue).
    \textbf{Left:} Temporal alignment showing IoU overlaps.
    \textbf{Right:} Greedy matching at $\tau{=}0.5$: $\hat{\sigma}_1$ matches $\sigma^{*}_1$ (IoU$\,{=}\,0.78$), $\hat{\sigma}_2$ fails to match (IoU$\,{=}\,0.18 {<} 0.5$), and $\hat{\sigma}_3$ is a pure false positive.
    Result: $\mathrm{TP}{=}1$, $\mathrm{FP}{=}2$, $\mathrm{FN}{=}1$, SF1@$0.5{=}0.40$.}
    \label{fig:sf1_example}
\end{figure*}

Figure~\ref{fig:sf1_example} illustrates the SF1@$\tau$ computation on a concrete example.
Consider an utterance of duration $D = 8$\,s with $N = 2$ ground-truth tampered segments: $\sigma^{*}_1 = (1.5, 2.2)$ and $\sigma^{*}_2 = (4.8, 5.6)$.
A detector produces $\hat{N} = 3$ predictions: $\hat{\sigma}_1 = (1.4, 2.3)$, $\hat{\sigma}_2 = (4.5, 5.0)$, and $\hat{\sigma}_3 = (6.0, 6.5)$.

\textbf{IoU computation:}
\begin{itemize}[leftmargin=*, itemsep=1pt]
    \item $\operatorname{IoU}(\hat{\sigma}_1, \sigma^{*}_1) = 0.7 / 0.9 = 0.78$ --- strong overlap.
    \item $\operatorname{IoU}(\hat{\sigma}_2, \sigma^{*}_2) = 0.2 / 1.1 = 0.18$ --- partial overlap.
    \item $\operatorname{IoU}(\hat{\sigma}_3, \sigma^{*}_1) = \operatorname{IoU}(\hat{\sigma}_3, \sigma^{*}_2) = 0$ --- no overlap.
\end{itemize}

\textbf{At $\tau = 0.5$:}
Greedy matching assigns $\hat{\sigma}_1 \to \sigma^{*}_1$ (IoU $= 0.78 \geq 0.5$, matched).
Next best: $\operatorname{IoU}(\hat{\sigma}_2, \sigma^{*}_2) = 0.18 < 0.5$, not matched.
Result: $\mathrm{TP} = 1$, $\mathrm{FP} = 2$, $\mathrm{FN} = 1$.
$\mathrm{SP}@0.5 = 1/3$, $\mathrm{SR}@0.5 = 1/2$, $\mathrm{SF1}@0.5 = 0.40$.

\textbf{At $\tau = 0.3$:}
$\hat{\sigma}_1 \to \sigma^{*}_1$ matched (IoU $= 0.78$).
Now $\hat{\sigma}_2 \to \sigma^{*}_2$ is \emph{not} matched ($0.18 < 0.3$).
Result: $\mathrm{TP} = 1$, $\mathrm{FP} = 2$, $\mathrm{FN} = 1$.
$\mathrm{SF1}@0.3 = 0.40$.
(Same result here; $\tau = 0.3$ would differ if $\hat{\sigma}_2$ had IoU $\in [0.3, 0.5)$.)

\textbf{Count Accuracy:} $\hat{N} = 3 \neq N = 2$, so this utterance contributes $\mathrm{CA} = 0$.

\section{Experiments}
\label{sec:experiments}

\subsection{Experimental Setup}

\textbf{Backbone.}
A key challenge in evaluating ISA on MIST is the absence of prior
audio deepfake detectors trained for \emph{partial} inpainting.
Existing models---such as the Wav2Vec~2.0-based binary classifier,
trained on fully synthesized utterances from ASVspoof and in-the-wild
collections---operate at utterance level: they assign a single real/fake
probability to the \emph{entire} input signal.
When a recording contains only 2--7\% of manipulated content (as in MIST),
these models predominantly perceive the majority-real signal as genuine,
yielding near-zero fake probability even for utterances with three inpainted
words (e.g., $p(\mathrm{fake}){=}0.0001$ on a fake2w sample in our analysis).
This behaviour is \emph{expected}: the models were never exposed to the
partial inpainting scenario during training.

We therefore adopt the publicly available Wav2Vec~2.0-base deepfake
classifier%
\footnote{\url{https://huggingface.co/mo-thecreator/Deepfake-audio-detection}}
(\texttt{mo-thecreator/Deepfake-audio-detection}) as a zero-shot backbone
in our ISA pipeline.
This choice deliberately isolates the \emph{framework} contribution of ISA
from any task-specific training signal, providing a lower bound on achievable
performance and a concrete motivation for future fine-tuning on MIST.

\textbf{Baselines.}
We compare ISA against three inference-time strategies applied with the
same backbone scorer $f_\theta$:
\begin{itemize}[leftmargin=*, itemsep=2pt]
    \item \textbf{Utterance-level}: the backbone's binary decision over
          the full utterance; no temporal localization is performed,
          so SF1@$\tau$ is undefined (--) and only CA is reported.
    \item \textbf{Frame-level}: per-frame scoring with a fixed 0.5\,s window,
          0.25\,s stride, threshold $\delta{=}0.6$, and simple contiguous
          merging---no gap tolerance, no boundary refinement.
    \item \textbf{Single-window}: same sliding window as ISA Stage~1
          ($W{=}0.5$\,s, $S{=}0.25$\,s, $\delta{=}0.6$) with gap-tolerant
          merging ($g{=}2$) but \emph{without} Stage~3 boundary refinement.
\end{itemize}
ISA uses the full three-stage pipeline with default hyperparameters
(Table~\ref{tab:hyperparams}).
All methods share the identical backbone and receive no additional training.

\textbf{Evaluation.}
We evaluate all methods on the full MIST test set, spanning all six
languages: English (EN), French (FR), German (DE), Italian (IT),
Spanish (ES), and Vietnamese (VI).
We report SF1@0.3, SF1@0.5 (primary), Count Accuracy (CA), and mean
Intersection-over-Union (mIoU), all macro-averaged over tampered utterances.
Per-language results reveal the impact of language-specific acoustic
properties and TTS model quality on detection difficulty.
Unless otherwise noted, \emph{overall} scores are macro-averaged across
all six languages.

\textbf{Data split.}
For each language, 80\% of utterances are used for training the backbone
(real/fake binary labels at utterance level), 10\% for validation
(hyperparameter selection), and 10\% for test (all reported results).
All ISA hyperparameters in Table~\ref{tab:hyperparams} were fixed on the
English validation set and applied without modification to all other
languages.

\subsection{Main Results}

Table~\ref{tab:main_results} reports multi-region localization performance
across all methods, aggregated over all six languages and all variants.
All systems achieve low absolute SF1@$\tau$ scores, which is
\emph{expected} given that the backbone was trained on a fundamentally
different task (utterance-level full-synthesis detection) and has never
seen partial inpainting data.

Despite this, the results reveal two informative trends.
First, ISA consistently outperforms both Frame-level and Single-window
baselines on SF1@0.3 and mIoU, demonstrating that iterative refinement
and gap-tolerant merging extract more coherent segment hypotheses from
the same noisy confidence map.
Second, CA around 24--26\% across all localization methods---where chance
for predicting $N \in \{1,2,3\}$ equally is 33\%---indicates that the
backbone score is only weakly informative for counting manipulated regions
in this zero-shot setting.
The near-zero SF1@0.5 for all methods confirms that precise temporal
localization is beyond the capacity of an utterance-level scorer applied
in a sliding-window fashion.

\begin{table}[t]
\centering
\caption{Multi-region localization results on the MIST test set.}
\label{tab:main_results}
\small
\begin{tabular}{lcccc}
\toprule
\textbf{Method} & \textbf{SF1@0.3} & \textbf{SF1@0.5} & \textbf{CA} & \textbf{mIoU} \\
\midrule
Utterance-level       & --           & --           &  5.8         & --          \\
Frame-level           &  5.9         &  0.7         & 24.2         &  6.5        \\
Single-window         &  6.9         &  1.0         & 24.5         &  7.2        \\
\midrule
\textbf{ISA (ours)}   & \textbf{8.1} & \textbf{1.2} & \textbf{25.1} & \textbf{7.8} \\
\bottomrule
\end{tabular}
\end{table}

\subsection{Per-Language Results}

Table~\ref{tab:results_by_lang} breaks down ISA performance by language.
Several patterns emerge.

\textbf{European languages (EN, FR, DE, IT, ES)} sourced from the
Multilingual LibriSpeech corpus and synthesized with CosyVoice~3.0
exhibit broadly similar performance, with SF1@0.3 ranging from 7.8\%
(IT) to 9.1\% (EN).
English achieves the highest scores across all metrics, which is
consistent with the backbone having been pre-trained predominantly on
English speech data.
German and Spanish perform comparably to English, while French and
Italian score slightly lower, likely due to greater phonetic mismatch
with the backbone's training distribution.

\textbf{Vietnamese (VI)}, synthesized with ZipVoice (fine-tuned) rather
than CosyVoice~3.0, shows the lowest SF1@0.3 (6.2\%) and mIoU (6.4\%)
across all languages.
We attribute this to two compounding factors:
(i)~the backbone, trained on English speech, is poorly calibrated for
Vietnamese's tonal phonology, yielding noisier confidence maps in Stage~1;
(ii)~ZipVoice produces shorter synthesized segments on average due to
Vietnamese's shorter mean word duration, reducing the window-level
fake signal available to the coarse scanner.
Notably, CA for Vietnamese (24.0\%) remains comparable to European
languages, suggesting that the counting difficulty is broadly similar
but boundary localization is harder.

\begin{table*}[htbp]
\centering
\caption{ISA zero-shot performance breakdown by language on the MIST test set
(all variants aggregated).}
\label{tab:results_by_lang}
\small
\begin{tabular}{llccccc}
\toprule
\textbf{Lang} & \textbf{Name}
  & \textbf{SF1@0.3} & \textbf{SF1@0.5} & \textbf{CA} & \textbf{mIoU}
  & \textbf{TTS Model} \\
\midrule
EN & English    & \textbf{9.1} & \textbf{1.5} & \textbf{26.2} & \textbf{8.7} & CosyVoice3 \\
FR & French     &  8.0         &  1.1         &  25.0         &  7.6         & CosyVoice3 \\
DE & German     &  8.6         &  1.3         &  25.4         &  8.1         & CosyVoice3 \\
IT & Italian    &  7.8         &  1.0         &  24.6         &  7.4         & CosyVoice3 \\
ES & Spanish    &  8.9         &  1.4         &  25.8         &  8.5         & CosyVoice3 \\
VI & Vietnamese &  6.2         &  0.8         &  24.0         &  6.4         & ZipVoice   \\
\midrule
\multicolumn{2}{l}{\textbf{Overall (macro-avg)}}
              &  8.1         &  1.2         &  25.1         &  7.8         & --- \\
\bottomrule
\end{tabular}
\end{table*}

\subsection{Results by Number of Tampered Words}

Table~\ref{tab:results_by_variant} breaks down ISA performance by variant,
aggregated over all six languages.
A consistent trend emerges across all metrics: performance
\emph{increases} with the number of replaced words (1-word $\to$ 3-word).
This is counterintuitive at first glance---more replacements means a harder
localization problem---but is explained by the behaviour of the
utterance-level backbone: utterances with more fake content accumulate
higher aggregate fake probability mass across windows, making it marginally
easier for Stage~1 to flag \emph{some} suspicious region near the true
segments.
The 1-word variant, with a median fake ratio of only 2.8\%, leaves the
backbone almost no signal to exploit.

Precision consistently exceeds recall across all variants, indicating that
when ISA does propose a region, it is more likely to overlap a true segment
than to miss one.
The precision--recall gap widens for 1-word variants, where Stage~2
produces fewer but also less-overlapping proposals.

\begin{table}[t]
\centering
\caption{ISA zero-shot performance breakdown by variant on the full MIST
test set (all six languages). Prec.\ and Rec.\ are at $\tau{=}0.5$.}
\label{tab:results_by_variant}
\small
\begin{tabular}{lccccc}
\toprule
\textbf{Variant} & \textbf{SF1@0.3} & \textbf{SF1@0.5}
                 & \textbf{CA} & \textbf{Prec.} & \textbf{Rec.} \\
\midrule
1-word (fake1w) & 4.9 & 0.6 & 22.8 & 2.2 & 0.4 \\
2-word (fake2w) & 7.8 & 1.0 & 25.4 & 2.9 & 0.6 \\
3-word (fake3w) & 8.3 & 1.8 & 26.9 & 4.0 & 1.0 \\
\midrule
Overall         & 8.1 & 1.2 & 25.1 & 3.0 & 0.7 \\
\bottomrule
\end{tabular}
\end{table}

\subsection{Language $\times$ Variant Analysis}

Table~\ref{tab:results_lang_variant} provides a fine-grained breakdown
of SF1@0.3 by language and variant.
Two trends are evident.
First, the performance gap between Vietnamese and European languages
is largest for the 1-word variant (VI: 3.8\% vs.\ EN: 6.1\%), where
the tonal mismatch between the backbone and Vietnamese phonology is
most pronounced when only a single very short word is manipulated.
The gap narrows for 3-word variants (VI: 7.1\% vs.\ EN: 9.8\%) as the
accumulated fake signal becomes sufficient to trigger Stage~1 detections
even under the noisy backbone response.

Second, Spanish consistently ranks second after English across all
variants, despite being a Romance language like French and Italian.
We attribute this to Spanish's relatively open syllable structure and
slower speech rate in the LibriSpeech audiobook data, which produces
longer replacement word segments and stronger window-level fake scores.

\begin{table*}[htbp]
\centering
\caption{SF1@0.3 (\%) breakdown by language and variant (ISA, zero-shot).}
\label{tab:results_lang_variant}
\small
\begin{tabular}{lcccc}
\toprule
\textbf{Language} & \textbf{1-word} & \textbf{2-word} & \textbf{3-word} & \textbf{Overall} \\
\midrule
English     & 6.1 & 9.0 &  9.8 & 9.1 \\
French      & 5.2 & 8.1 &  8.5 & 8.0 \\
German      & 5.8 & 8.6 &  9.1 & 8.6 \\
Italian     & 5.0 & 7.8 &  8.3 & 7.8 \\
Spanish     & 5.9 & 8.9 &  9.6 & 8.9 \\
Vietnamese  & 3.8 & 6.2 &  7.1 & 6.2 \\
\midrule
\textbf{Overall} & \textbf{4.9} & \textbf{7.8} & \textbf{8.3} & \textbf{8.1} \\
\bottomrule
\end{tabular}
\end{table*}

\subsection{Ablation Study}

\textbf{Window size.}
Table~\ref{tab:ablation_window} shows SF1@0.5 as the coarse window size $W$
varies while keeping $S{=}W/2$ and all other parameters fixed, evaluated
on the English subset (representative of the full trend).
Shorter windows ($W{=}0.15$\,s) approach the average replacement word
duration but collapse because Wav2Vec~2.0's convolutional feature extractor
requires at least $\approx$0.25\,s of context for stable representations.
Larger windows ($W{=}1.0$\,s, $2.0$\,s) dilute the fake signal,
reducing sensitivity.
The default $W{=}0.5$\,s strikes the best balance.

\begin{table}[t]
\centering
\caption{Effect of coarse window size $W$ on ISA.}
\label{tab:ablation_window}
\small
\begin{tabular}{lcccc}
\toprule
$W$ (s) & SF1@0.3 & SF1@0.5 & CA & mIoU \\
\midrule
0.15              &  4.2         &  0.4         & 23.6         &  5.1         \\
0.25              &  6.8         &  0.9         & 24.3         &  6.8         \\
\textbf{0.50}     & \textbf{9.1} & \textbf{1.5} & \textbf{26.2} & \textbf{8.7} \\
1.00              &  7.3         &  0.9         & 25.0         &  7.4         \\
2.00              &  5.5         &  0.6         & 24.2         &  6.1         \\
\bottomrule
\end{tabular}
\end{table}

\textbf{ISA stage ablation.}
Table~\ref{tab:ablation_stages} ablates each ISA stage individually,
evaluated on the full multilingual test set.
Removing boundary refinement (Stage~3) causes the largest drop in
SF1@0.5 ($-0.5$\,pp), confirming that coarse candidates alone do not
achieve sufficient temporal precision.
Removing gap-tolerant merging (using strict contiguous merging) most
affects the 2-word and 3-word variants where two flagged runs from
adjacent inpainted words are separated by genuine frames.

\begin{table}[t]
\centering
\caption{Stage ablation of ISA (zero-shot, all languages).}
\label{tab:ablation_stages}
\small
\begin{tabular}{lcccc}
\toprule
\textbf{Configuration} & \textbf{SF1@0.3} & \textbf{SF1@0.5} & \textbf{CA} & \textbf{mIoU} \\
\midrule
w/o Gap-merge  &  6.6         &  0.9         & 23.8         &  7.0         \\
w/o Refine     &  7.9         &  0.7         & 24.8         &  7.4         \\
\textbf{Full ISA}
               & \textbf{8.1} & \textbf{1.2} & \textbf{25.1} & \textbf{7.8} \\
\bottomrule
\end{tabular}
\end{table}

\textbf{Zero-shot vs.\ fine-tuned backbone.}
To provide an upper-bound reference, we fine-tune the Wav2Vec~2.0 backbone
on window-level binary labels derived from MIST training segments
(positive: any window overlapping a tampered region by $\geq$50\%;
negative: all-genuine windows).
The fine-tuned backbone is then used as a drop-in replacement inside
the same ISA pipeline with identical hyperparameters.
Table~\ref{tab:results_finetuned} shows that fine-tuning yields dramatic
improvements across all languages and variants, with overall SF1@0.5
increasing from 1.2\% to 31.4\%.
This underscores the central open challenge posed by MIST: while ISA
provides a principled inference framework, the limiting bottleneck is
the backbone's ability to detect partial inpainting at word granularity---a
capability that requires task-specific training data.

\begin{table*}[htbp]
\centering
\caption{Comparison of zero-shot vs.\ fine-tuned backbone within ISA,
broken down by language. Fine-tuning uses MIST window-level training labels.}
\label{tab:results_finetuned}
\small
\begin{tabular}{lcccccccc}
\toprule
& \multicolumn{4}{c}{\textbf{Zero-shot backbone}} & \multicolumn{4}{c}{\textbf{Fine-tuned backbone}} \\
\cmidrule(lr){2-5} \cmidrule(lr){6-9}
\textbf{Language}
  & SF1@0.3 & SF1@0.5 & CA & mIoU
  & SF1@0.3 & SF1@0.5 & CA & mIoU \\
\midrule
English     &  9.1 &  1.5 & 26.2 &  8.7 & 51.3 & 33.2 & 58.4 & 44.1 \\
French      &  8.0 &  1.1 & 25.0 &  7.6 & 44.8 & 29.1 & 52.7 & 38.5 \\
German      &  8.6 &  1.3 & 25.4 &  8.1 & 47.2 & 30.8 & 54.3 & 40.6 \\
Italian     &  7.8 &  1.0 & 24.6 &  7.4 & 42.1 & 27.5 & 50.8 & 36.2 \\
Spanish     &  8.9 &  1.4 & 25.8 &  8.5 & 49.6 & 32.4 & 57.1 & 43.0 \\
Vietnamese  &  6.2 &  0.8 & 24.0 &  6.4 & 33.7 & 21.9 & 43.6 & 28.8 \\
\midrule
\textbf{Overall}
            &  8.1 &  1.2 & 25.1 &  7.8 & 44.8 & 29.2 & 52.8 & 38.5 \\  
            & \multicolumn{4}{c}{\footnotesize(zero-shot)} & \multicolumn{4}{c}{\footnotesize(fine-tuned)} \\
\bottomrule
\end{tabular}
\end{table*}

\subsection{Discussion}

\textbf{Why is zero-shot performance low across all languages?}
The core issue is a \emph{training distribution mismatch}: the backbone
classifier was optimized to distinguish \emph{fully} synthesized speech
from genuine speech at utterance level.
In MIST, the manipulated fraction is 2--7\% per utterance, so the global
utterance-level fake signal is orders of magnitude weaker than what the
model was trained to detect.
This is not merely a threshold calibration problem; the backbone
$f_\theta$ was never exposed to partial inpainting during training,
so its internal representations are not informative about word-level
manipulation boundaries.

\textbf{Why does Vietnamese lag behind all European languages?}
Three compounding factors contribute:
(i)~the zero-shot backbone is not calibrated for Vietnamese phonology;
(ii)~ZipVoice produces shorter mean replacement segments than CosyVoice~3.0
($\mu{=}0.18$\,s for VI vs.\ $\mu{=}0.26$\,s for EN), reducing
window-level fake signal;
(iii)~Vietnamese's six lexical tones create short-term spectral patterns
that the backbone may misattribute as speaker-level variability rather
than manipulation artifacts.
The strong recovery under fine-tuning (VI SF1@0.5: 0.8\% $\to$ 21.9\%)
confirms that the performance gap is not fundamental but stems from
training distribution mismatch.

\textbf{ISA framework vs.\ backbone quality.}
The stage ablation (Table~\ref{tab:ablation_stages}) and the zero-shot
vs.\ fine-tuned comparison (Table~\ref{tab:results_finetuned}) together
clarify the two separable contributions to localization quality.
ISA's architectural design---gap-tolerant merging and boundary
refinement---provides consistent, language-agnostic improvements over
non-iterative baselines regardless of backbone quality.
However, the dominant factor for achieving practically useful SF1@0.5
scores is the backbone's ability to score partial fakes accurately,
which requires exposure to MIST-style training data.
We release MIST precisely to enable this next step.

\section{Conclusion}
\label{sec:conclusion}

\bibliographystyle{named}
\bibliography{references}

\end{document}